\documentclass[%
APS,
rsi,%
 amsmath,amssymb,superscriptaddress,
reprint,
]{revtex4-2}
\usepackage{lipsum}
\usepackage{subfigure}
\usepackage{graphicx}
\usepackage{dcolumn}
\usepackage{bm}
\usepackage{siunitx}
\usepackage{xcolor}
\usepackage{gensymb}
\usepackage{ulem}
\usepackage[export]{adjustbox}

\begin{document}

\preprint{AIP/123-QED}

\title[Sample title]{Effect of multi-layering and crystal orientation on spin-orbit torque efficiency in Ni/Pt layer stacking}

\author{A.~Sud}
\email{aakanksha.sud.c1@tohoku.ac.jp}
 \affiliation{WPI-AIMR, Tohoku University, 2-1-1, Katahira, Sendai 980-8577, Japan}
 \affiliation{London Centre for Nanotechnology, University College London, 17-19 Gordon Street, London, WC1H 0AH, United Kingdom}
 \author{Y.-C.~Lau}%
 \affiliation{Institute for Materials Research, Tohoku University, 2-1-1, Katahira, Sendai 980-8577, Japan}
 \affiliation{ Institute of Physics, Chinese Academy of Sciences, Beijing 100190, China}
 \author{J.~Brierley}
\affiliation{London Centre for Nanotechnology, University College London, 17-19 Gordon Street, London, WC1H 0AH, United Kingdom}
\affiliation{Department of Electronic $\And$ Electrical Engineering, University College London, 7JE, Malet Pl, London WC1E 7JE}
\affiliation{ISIS Neutron and Muon Source, Rutherford Appleton Laboratory, Harwell Oxford, Didcot OX11 0QX}
\author{H.~Kurebayashi}
\affiliation{London Centre for Nanotechnology, University College London, 17-19 Gordon Street, London, WC1H 0AH, United Kingdom}
\affiliation{Department of Electronic $\And$ Electrical Engineering, University College London, 7JE, Malet Pl, London WC1E 7JE}
 \affiliation{WPI-AIMR, Tohoku University, 2-1-1, Katahira, Sendai 980-8577, Japan}
\author{T.~Seki}%
 \email{takeshi.seki@tohoku.ac.jp}
 \affiliation{Institute for Materials Research, Tohoku University, 2-1-1, Katahira, Sendai 980-8577, Japan}


\date{\today}

\begin{abstract}
 We study spin-orbit torques (SOTs) in Ni/Pt bi-layers and multi-layers by ferromagnetic resonance (FMR) and harmonic-Hall measurements. The effect of multi-layering and crystal orientation on field-like (FL) and damping-like (DL) torque efficiencies is examined by exploiting the  samples with different crystal orientations: epitaxial and poly-crystalline structures on Sapphire and SiO$_2$ substrates, respectively. 
 We find that both DL and FL torque efficiencies are larger in multi-layer samples and there is no complete cancellation of torque efficiencies that is generally expected for ideal symmetric stacking structures. The results of SOT-FMR indicate that the epitaxial samples show higher efficiency for SOT generation compared to the poly-crystalline samples, suggesting that SOT generation is modified depending on the interfacial contribution. In addition, the spin Hall conductivity of the epitaxial multi-layer is the largest among the samples. The present results signify the importance of crystal orientation, multi-layering and interface-quality in improving the efficiency of SOTs generation combined with larger spin hall angle for developing future spintronic devices. 
\end{abstract}

 \maketitle
\section{\label{sec:level1}Introduction\protect\\}
Spin-orbit torques (SOTs) \cite{liu2011spin,manchon2019current,miron2011perpendicular,kurebayashi2014antidamping,nakayama2016rashba,brataas2012current,hayashi2014quantitative,garello2013symmetry}  provide an efficient way to manipulate the magnetisation in ferromagnetic single layers or hetero-structures with large spin-orbit coupling. 
Two mechanisms responsible for the SOTs in ferromagnetic/heavy-metal (FM/HM) bi-layers are based on creation of non-equilibrium spin polarisation by electric current \cite{manchon2019current}. The first mechanism is  attributed to the spin Hall effect (SHE) \cite{hirsch1999spin,sinova2004universal,kato2004observation} which occurs in a non-magnetic layer that converts a charge current into a transverse spin current. The generated spin-current then diffuses to the ferromagnetic layer through spin angular momentum transfer which exerts a torque. The second mechanism of generation of SOTs occurs at the interface due to interfacial spin-orbit coupling in the FM/HM bi-layers. When the inversion symmetry is broken at the interface, the spin-orbit hamiltonian lifts the degeneracy of electron-spin momentum states. Because of this spin-orbit term, a charge current flowing parallel to the interface is to be spin-polarized. This process is called Rashba-Edelstein effect (Inverse spin galvanic effect) \cite{bychkov1984oscillatory,edelstein1990spin,sanchez2013spin}. The resultant spin polarisation can exert torques in the adjacent FM layer \cite{jungfleisch2016interface}. Both mechanisms can result in damping-like (DL) torque $\sim \tau_{\textrm {DL}}(\mathbf{m}\times(\mathbf{\sigma}\times\mathbf{m}))$ and field-like (FL) torque $\sim \tau_{\textrm {FL}}(\mathbf{\sigma}\times\mathbf{m})$ \cite{PhysRevB.87.174411}. Here, $\mathbf{m}$ is the unit vector of magnetisation, $\mathbf{\sigma}$ is the spin polarization in direction of spin-current and $\tau_{\textrm {DL(FL)}}$ is the magnitude of torque which has already been described in details in the previous  works \cite{kim2012magnetization,abert2017fieldlike,dutta2021interfacial}. In real systems it might be intuitively appealing to track down and parse the resultant torque into these two contributions. However, SOT scenario is often complicated due to the presence of local spin currents \cite{wang2019anomalous} within the ferromagnet which can also contribute to torque generation. Moreover, in thin films the existence of orbital effects can produce unconventional torques \cite{hayashi2023observation}. Understanding the interplay between bulk, interfacial and orbital contributions to SOT is important to enhance the efficiency of SOT devices.

The SOT efficiency is a figure of merit for characterizing its potential for future spintronic applications. New materials which can offer higher SOT efficiency are continuously being investigated including ferromagnets, anti-ferromagnets \cite{zhou2019large}, 2-dimensional materials \cite{ahn20202d}, heavy metals \cite{choi2020optical} and topological insulators \cite{han2021topological}. The quest for efficient sources for SOT generation raises many questions about the mechanism and the magnitude of their effectiveness in SOT enhancement. What is the nature of spin orbit effects that can lead to larger SOTs? Is there any significant variation in torque efficiency dependent on multi-layer composition, thickness or interface modification? A number of such open questions remain that can have impact on future developments.  A proper interpretation of torque efficiency is highly complex owing to its origin in both bulk as well as interfacial effects \cite{hayashi2021spin}. The interface quality as well as film structure can have prolific influence on the torque efficiency \cite{PhysRevB.91.214416, lee2019enhanced, an2018giant}, shedding light on the importance of sample growth conditions that significantly influence the SOT efficiency \cite{berger2018determination, pai2015dependence, zhang2015role}. Moreover, spin relaxation mechanisms are dependent on crystal morphology and structure which can influence SOTs \cite{long2016strong, freeman2018evidence}. Both Elliot-Yafet (EY) \cite{elliott1954theory,yafet1952calculation} and Dyakanov-Perel (DP) relaxation \cite{long2016strong} mechanisms can be modulated by structural changes \cite{ryu2016observation}. Additionally, enhanced SOT efficiencies have been reported in multi-layers with ferromagnet sandwiched between two non-magnets having opposite spin Hall angle \cite{woo2014enhanced}. The presence of planar Hall effect causes additional anti-damping torque in asymmetric structures. These facts reported by previous studies imply that the crystal structure \cite{ryu2019crystal, choi2022thickness, kumar2020impact} as well as the interfacial effects play a role in SOTs generation and modification. 

In this paper we study the effect of multi-layering and crystal orientation on the generation of SOTs in Ni/Pt bi-layers and multi-layers. This is done using ferromagnetic resonance (FMR) and harmonic-Hall measurements.  There have been some interesting results on SOT engineering in multi-layer systems for example in Co/Pt multi-layers \cite{huang2015engineering,wu2019enhanced,ishikuro2020highly} and some other works demonstrating the role played by interfaces in the origin of unconventional SOT behaviours in multi-layer  stacks of  Co/Pd \cite{jamali2013spin}. These works suggest that multi-layers will provide a research stage different from the bi-layer cases. Apart from several works in Co based multi-layers, there has been only one work investigating SOT in Ni/Pt bi-layer system \cite{hayashi2021spin}. However, they do not report the case of multi-layer. Both Co and Ni have face-centred cubic lattices, but very different electronic structures giving rise to totally different magnetic and transport properties. This means that Ni/Pt is a candidate system to do the extensive study on the potential of multi-layer as a SOT material. Since epitaxial growth is also possible for the Ni/Pt multi-layers, Ni/Pt is considered to be a suitable material to investigate the influence of film crystallinity on the magnitude of SOT. This should provide promising directions towards the development of spin-orbit torque in other thin-film stacks. The samples used for this study are grown on sapphire and SiO$_2$ substrates which show epitaxial and poly-crystalline growth as revealed by our structural analysis. We found DL torque efficiency, from FMR measurements, shows a large enhancement in multi-layer epitaxial samples which is about 160$\%$ larger than bi-layer poly-crystalline samples. The values for DL torque efficiency are in agreement with those obtained from harmonic-hall measurements. The multi-layer epitaxial samples also exhibit largest FL torque efficiencies. Furthermore, the results of FMR linewidth modulation confirm larger spin Hall angle $\sim$ 0.15 in these samples. Our results demonstrate the importance of growth on SOT efficiency and also its effect on various magnetic parameters. We illustrate an efficient approach to improve the SOT efficiency, which is achieved by growing multi-layer epitaxial samples. 

\section{Sample Growth, Characterisation and device fabrication}
\subsection{\label{sec:level2}Film growth}
The Ni/Pt multi-layer films were grown using magnetron sputtering in an ultrahigh vacuum system with a base pressure below $2\times10^{-7}$ Pa. Two types of substrates were used - $\textrm{sapphire}(0001)$ and $\textrm{SiO}_2$ for epitaxial and polycrystalline growth, respectively. A Pt layer was first deposited on top of the substrate followed by growth of the Ni layer. The thicknesses of Ni and Pt layer were fixed at 3 and 1 nm, respectively, for all the samples. The sample label and stack structure is summarized in Table.~\ref{tab:sample} where $\textrm{[Ni/Pt]}_{\times \textrm{5}}$ represents the five time repetition of $\textrm{[Ni/Pt]}$ layers.
\begin{table*}
\centering
\small
 \caption{\label{tab:sample} Summary of the sample notation used, stacking pattern, deposition temperature and substrate used. The RHEED and XRD confirm the epitaxial/polycrystalline growth. The numbers in brackets represent thicknesses in nm. The easy axis of magnetisation are also presented.}
\begin{tabular}{cccccc}
\hline\hline
Sample&  Substrate &Deposition & Stack &RHEED, &Easy  \\
  &  &Temperature & &XRD &Axis\\
\hline
K-01 &Sapphire &400$^{\degree}$C &[Pt (1)/Ni(3)]$_{\times 5}\vert \textrm{Al}_2\textrm{O}_3 (5)$ &(111) epi. &In-Plane \\
K-02 &SiO$_2$ &Room Temp. &[Pt (1)/Ni(3)]$_{\times 5}\vert\textrm{Al}_2\textrm{O}_3 (5)$ &Poly. &In-Plane\\
K-03 &Sapphire  &400$^{\degree}$C  &[Pt (1)/Ni(3)]$\vert\textrm{Al}_2\textrm{O}_3 (5)$ &(111) epi. &In-Plane\\
K-04 &SiO$_2$ &Room Temp. &[Pt (1)/Ni(3)]$\vert\textrm{Al}_2\textrm{O}_3 (5)$ &Poly. &In-Plane\\
\hline
\hline
\end{tabular}
\end{table*}

The deposition temperature was optimized to achieve the epitaxial growth on sapphire and to ensure successful formation of the layered structure \cite{seki2020perpendicularly}. These conditions were met at a temperature of $400^{\degree}$ C which was maintained constant throughout the deposition process. Finally, a 5 nm thick $\textrm{Al}_2\textrm{O}_3$ capping layer was deposited on the top. The magnetic anisotropy induced at the interface of Ni/$\textrm{Al}_2\textrm{O}_3$ was negligible, confirmed by using a reference sample of Ni single layer film with $\textrm{Al}_2\textrm{O}_3$ capping layer. The samples K-02 and K-04 were deposited  on $\textrm{SiO}_2$ substrate at room temperature.

\subsection{\label{sec:level3}Film characterisation}
\begin{figure}
\centering
{\includegraphics[width=0.35\textwidth]{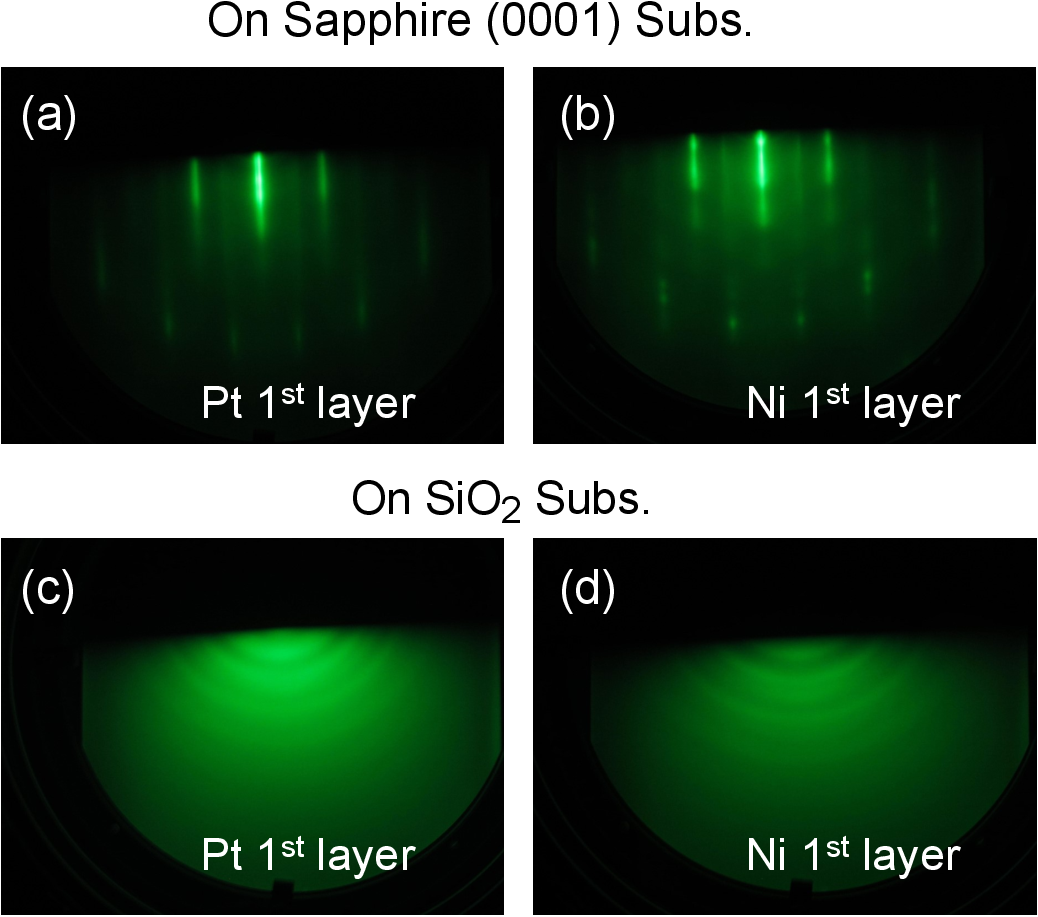}}
\caption{\label{fig:ch3_rheed} Reflection high-energy electron diffraction patterns for samples K-01 and K-02 grown on sapphire and SiO$_2$, respectively. The diffraction patterns were observed just after the growth of the first Ni and Pt layers.}
\end{figure}
\begin{figure}
\includegraphics[width=0.47\textwidth]{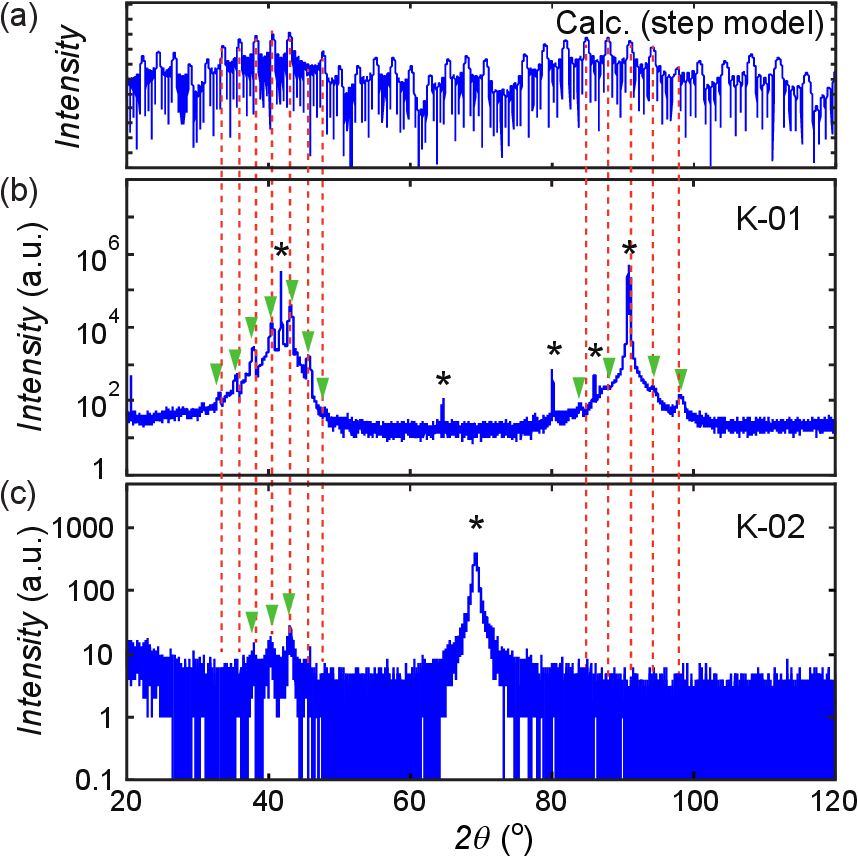}
\caption{\label{fig:ch3_xrd}X-ray diffraction profiles of (a) calculation, (b) K-01 sample grown at $400\degree$ C on sapphire substrate and (c) K-02 sample grown at room temperature on $\textrm{SiO}_2$ substrate. The black asterisks denote the diffractions from the substrates and the green inverted triangles denote the diffractions of multi-layered structures. The calculation was done using the Laue function with the assumption of the step model for the superstructure. The red dotted lines are guides indicating the peak positions for the superlattice with the (111) crystal orientation. }
\end{figure}

\begin{figure}
\centering
{\includegraphics[width=0.45\textwidth]{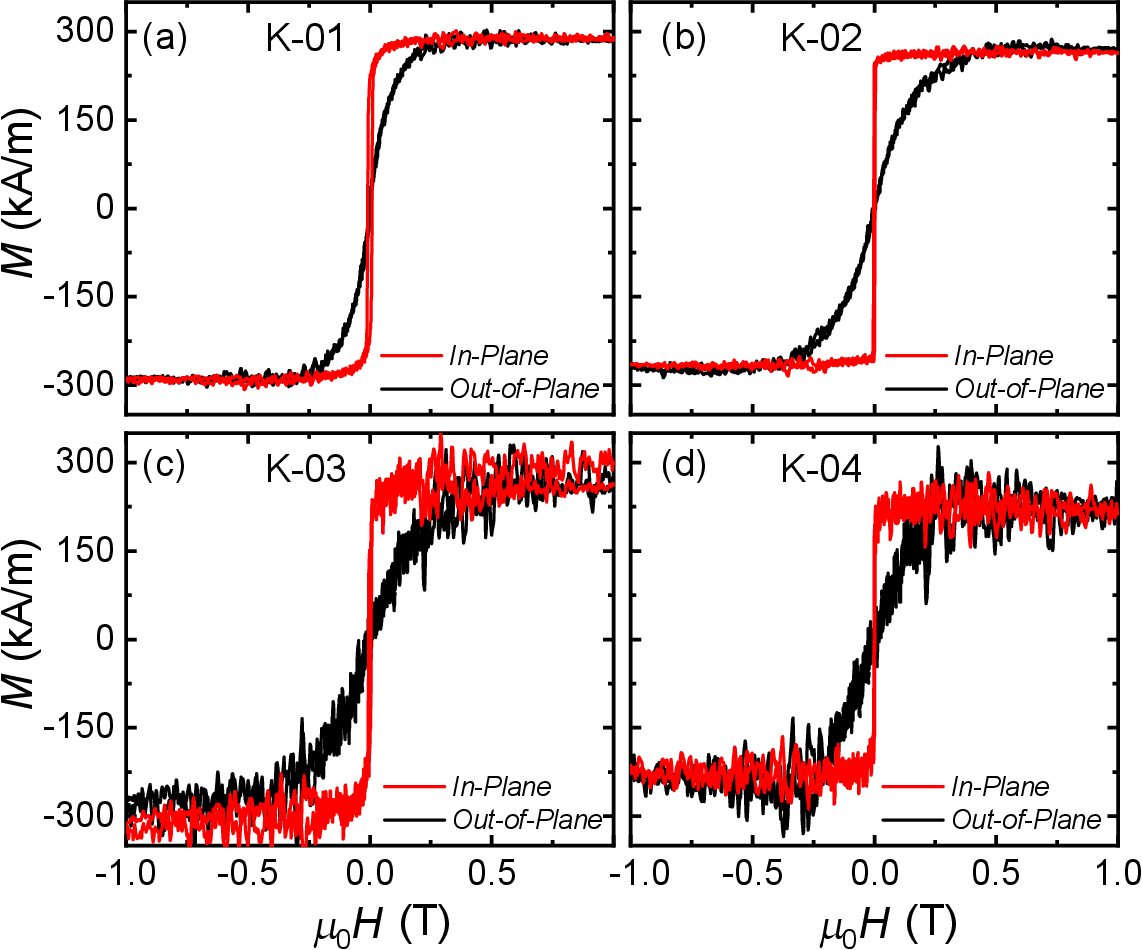}}
\caption{\label{fig:ch3_vsm}Magnetisation curves for the samples measured using vibrating sample magnetometer (VSM). The red curves are the results measured with in-plane magnetic field (IP) while the black curves denote the measurements with out-of-plane magnetic field (OOP). The measurements were done at room temperature.}
\end{figure}

The crystal orientation and morphology were monitored $in\textendash situ$ during growth using reflection high-energy electron diffraction (RHEED) and we present some of them in Fig.~\ref{fig:ch3_rheed}. Polycrystalline growth is confirmed by the presence of rings for the sample grown on a $\textrm{SiO}_2$ substrate (Figs.~\ref{fig:ch3_rheed}(c) and 1(d)) while sharp periodic streaks correspond to epitaxial growth on a sapphire substrate (Figs.~\ref{fig:ch3_rheed}(a) and 1(b)). Structural characterisation was performed using x-ray diffraction (XRD) techniques with Cu-$K\alpha$ radiation shown in Fig.~\ref{fig:ch3_xrd} for K-01 and K-02. The diffraction profile calculated by the Laue function with the assumption of the step model for the superstructure is also displayed on top of the XRD profiles. For this model $\textrm{[Ni (3 nm) /Pt (1 nm) ]}_{\times \textrm{5}}$ with the (111) crystal orientation was assumed (See Appendix for more details). The XRD profile of K-01 shows a clear multiple-peak structure, whose peak positions are explained by the calculated XRD profiles, indicating that the K-01 film is a well-defined superlattice with the designed thicknesses and the (111) crystal orientation. In contrast to K-01, sample K-02 exhibits a few small peaks around 2$\theta$ = 40$\degree$. Taking into account the fact that the RHEED patterns are the ring shapes for K-02, we confirm that K-02 is a polycrystalline film. However, its XRD peak positions fairly match with those calculated. This suggests that the (111) crystal orientation is preferential in the film, despite its polycrystalline behaviour confirmed by RHEED. An important point is that the calculated XRD profile uses the bulk lattice constant of Ni and Pt, and the calculation results can explain the peak positions of both K-01 and K-02, suggesting that there is no remarkable difference in the lattice constant between K-01 and K-02. Thus, from both RHEED and XRD results it is confirmed that the samples grown on sapphire substrates show epitaxial growth whereas those on SiO$_2$ form a polycrystalline film.
 
The magnetic hysteresis loops were measured using a vibrating sample magnetometer (VSM) at room temperature. Figure~\ref{fig:ch3_vsm} shows the magnetisation curves for the four samples obtained when applying field $\mu_0H$ along  in-plane (red curves) and out-of-plane directions (black curves) of the films. All the samples are in-plane favoured materials with coercivity of approximately 10 mT. Table~\ref{tab:sot2} summarizes the estimated saturation magnetisation values ($\mu_0 M_{\textrm{s}}$) for the different samples.
\subsection{\label{sec:level4}Device fabrication}
In order to study the spin transport by SOT-FMR measurements, the samples were cut into 5 mm $\times$ 5 mm chips prior to device fabrication using standard lithography and Ar-ion milling techniques. Rectangular bars of width 5 $\mu$m and length of 400 $\mu$m (K-01 and K-02) and 200 $\mu$m (K-03 and K-04) were defined for FMR measurements. The device circuitry and bar pattern are illustrated in Fig.~\ref{fig:ch3_schema}(a). For harmonic-Hall resistance measurements, the samples were patterned in the form of symmetric Hall cross structures as shown in the inset of Fig.~\ref{fig:ch3_harmonic} (a). The length and width of the Hall bar were 25 $\mu$m and 10 $\mu$m respectively.
\section{Measurement Techniques}
We quantified the SOT efficiency and magnetic parameters using two independent techniques of SOT-FMR and harmonic-Hall resistance measurements as described below. 
\subsection{Spin-orbit torque Ferromagnetic Resonance}
\begin{figure}
\centering
{\includegraphics[width=0.48\textwidth]{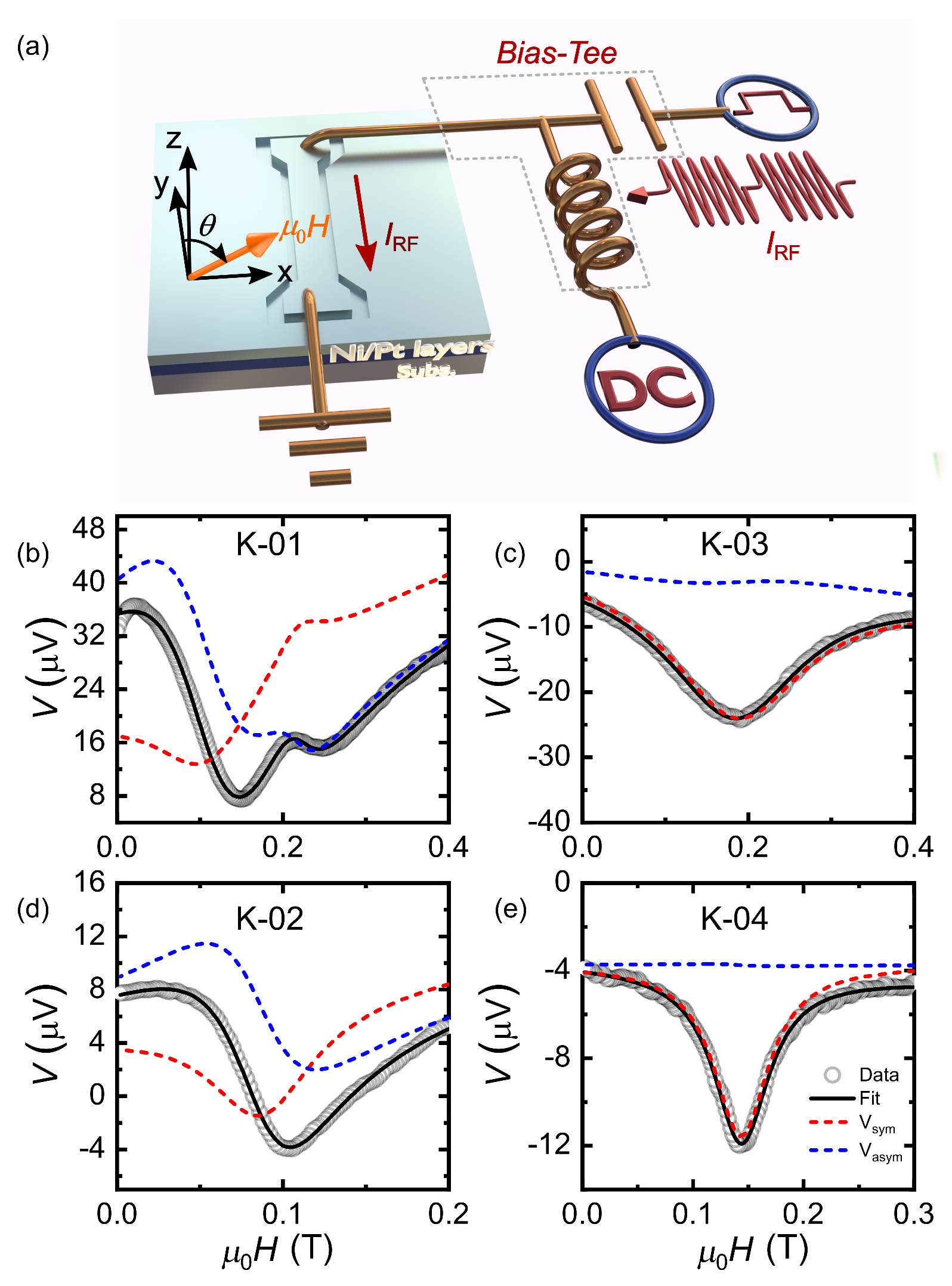}}
\caption{\label{fig:ch3_schema}Schematic illustration of the experimental setup used for SOT-FMR experiments. The microwave current, $I_{\textrm{RF}}$ is injected into the bar with width 5 $\mu m$ through a bias-tee. $\theta$ denotes the in-plane angle between the external field $\mu_0H$ and the microstrip. (b-d) Magnetic field dependence of the DC voltage, $V$, for the samples measured at 8 GHz. The data points are fitted (solid black line) with a combination of symmetric (dashed red line)  and anti-symmetric (dashed blue line). A clear anti-symmetric component is visible for multi-layer sample as compared to bi-layer sample (K-03, K-04). }
\end{figure}

By FMR measurements, we are able to extract the SOT efficiency in Ni/Pt multi-layer films. The measurement setup is illustrated in Fig.~\ref{fig:ch3_schema}(a). Microwaves at a fixed frequency, $f$, were injected into the bar while sweeping the external dc magnetic field at different angles, $\theta$, from the bar direction. The effective magnetic fields $(h_{\textrm{x}},h_{\textrm {y}},h_{\textrm {z}})$ are generated from an injected microwave current, which then exert SOT on the adjacent FM layer into precession. This magnetisation precession causes a time varying change in  sample resistance due to anisotropic magnetoresistance (AMR) and produces a DC voltage, $V$, due to rectification \cite{kurebayashi2014antidamping}. Figure~\ref{fig:ch3_schema} (b-e) show a typical voltage signal for different samples measured at $f = 8$ GHz and $\theta = 45\degree$ together with the fitted curves obtained using the sum of symmetric and anti-symmetric Lorentzian given by Eq.~\ref{eq:voltage} (see Appendix D for details).
\begin{figure}
\centering
{\includegraphics[width=0.47\textwidth]{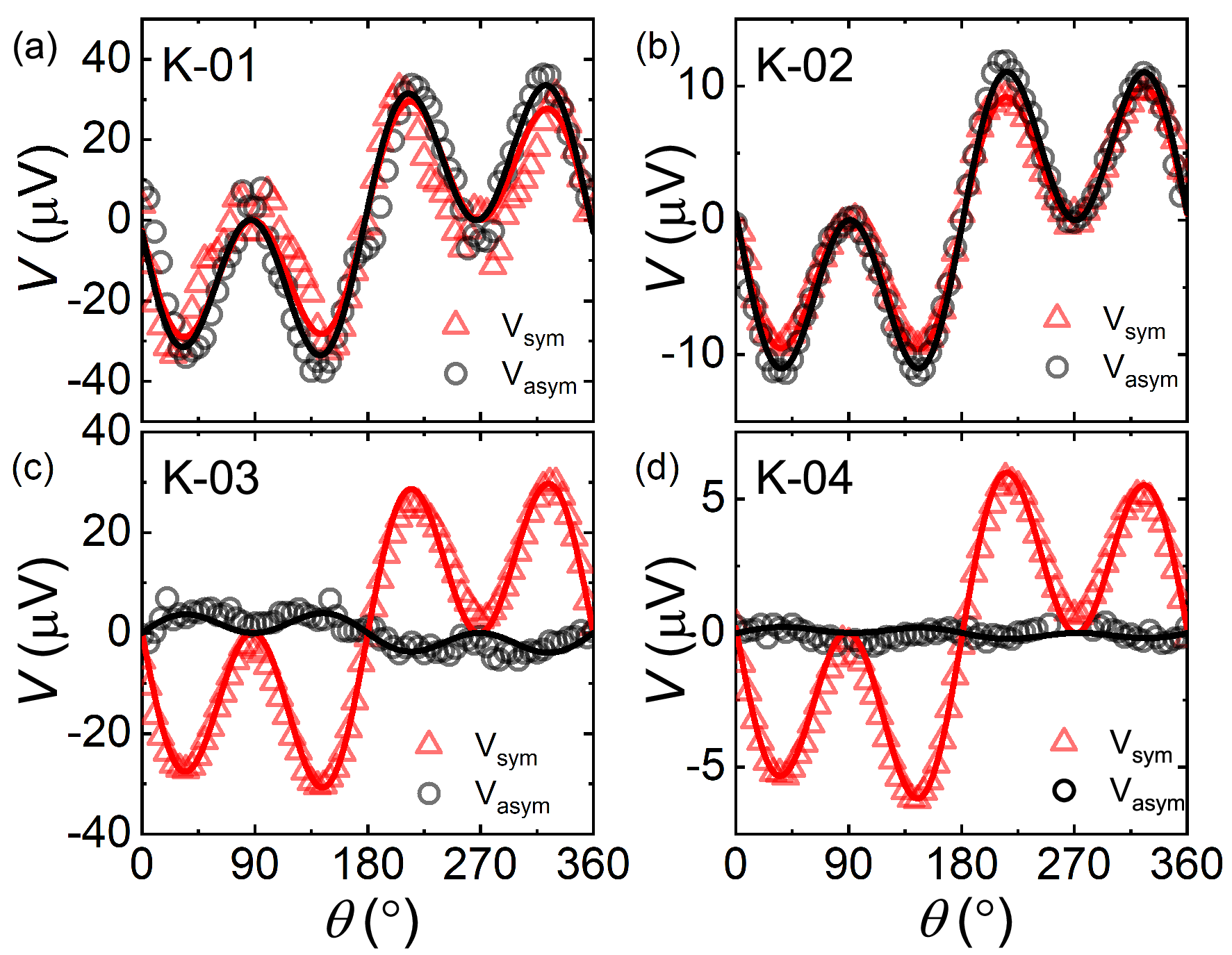}}
\caption{\label{fig:ch3_sot}The symmetric and anti-symmetric components of the SOT-FMR spectra as a function of in-plane magnetic field angle $\theta$ for (a) K-01 (b) K-02 (c) K-03 and (d) K-04 films at $f=$ 8 GHz. The bi-layer samples show negligible anti-symmetric component while the anti-symmetric component is comparable to symmetric component for the multi-layer samples.}
\end{figure}

The DL and FL SOT efficiencies are calculated using the value of effective fields obtained by fitting angular dependence of symmetric and anti-symmetric components of $V$ (see Appendix D). The symmetric component of voltage, $V_\textrm{sym}$ $\propto h_{\textrm {z}}$, is related to DL torques whereas the FL torques and Oersted torques are obtained from the anti-symmetric component, $V_\textrm{asym}$ $\propto h_{\textrm {x}}$, $ h_{\textrm {y}}$. In Fig.~\ref{fig:ch3_sot} we plot the $V_\textrm{sym}$ and $V_\textrm{asym}$ obtained at different $\theta$ for all samples. The  black (red) solid lines shown in Fig.~\ref{fig:ch3_sot} are obtained using Eq.~\ref{eq:F14} (\ref{eq:F13}) from which we extract the values of $h_{\textrm x},h_{\textrm {y}}$ ($h_{\textrm {z}}$) summarised in Table.~\ref{tab:sot}. We allowed $h_{\textrm {z}}$ in Eq.~\ref{eq:F13} to be angular-dependent as $h_{\textrm {z}} =a +b\cos\theta+c\sin\theta$. The field values are summarised in Table~\ref{tab:sot} in Appendix. $h_\textrm{y}$ and $b$ are dominant in FL and DL torques respectively, consistent with the standard Oersted FL and spin-hall DL pictures. From these values we obtain the combined FL spin-orbit and Oersted effective field  ($H_{\textrm{FL+Oe}}$) and DL spin-orbit effective field ($H_\textrm{DL}$ ) respectively. It can be seen from Fig.~\ref{fig:ch3_sot} that bi-layer films have very weak value of $V_\textrm{asym}$ component. Interestingly, the sign of $V_\textrm{sym}$ is constant in all samples whereas the sign of $V_\textrm{asym}$ flips in the bi-layer films. This suggests the presence of an additional FL torque term other than the Oersted origin, which seems to be thickness and material parameter dependent \cite{skinner2014spin} in our bi-layers.

\begin{figure}
\centering
{\includegraphics[width=0.48\textwidth]{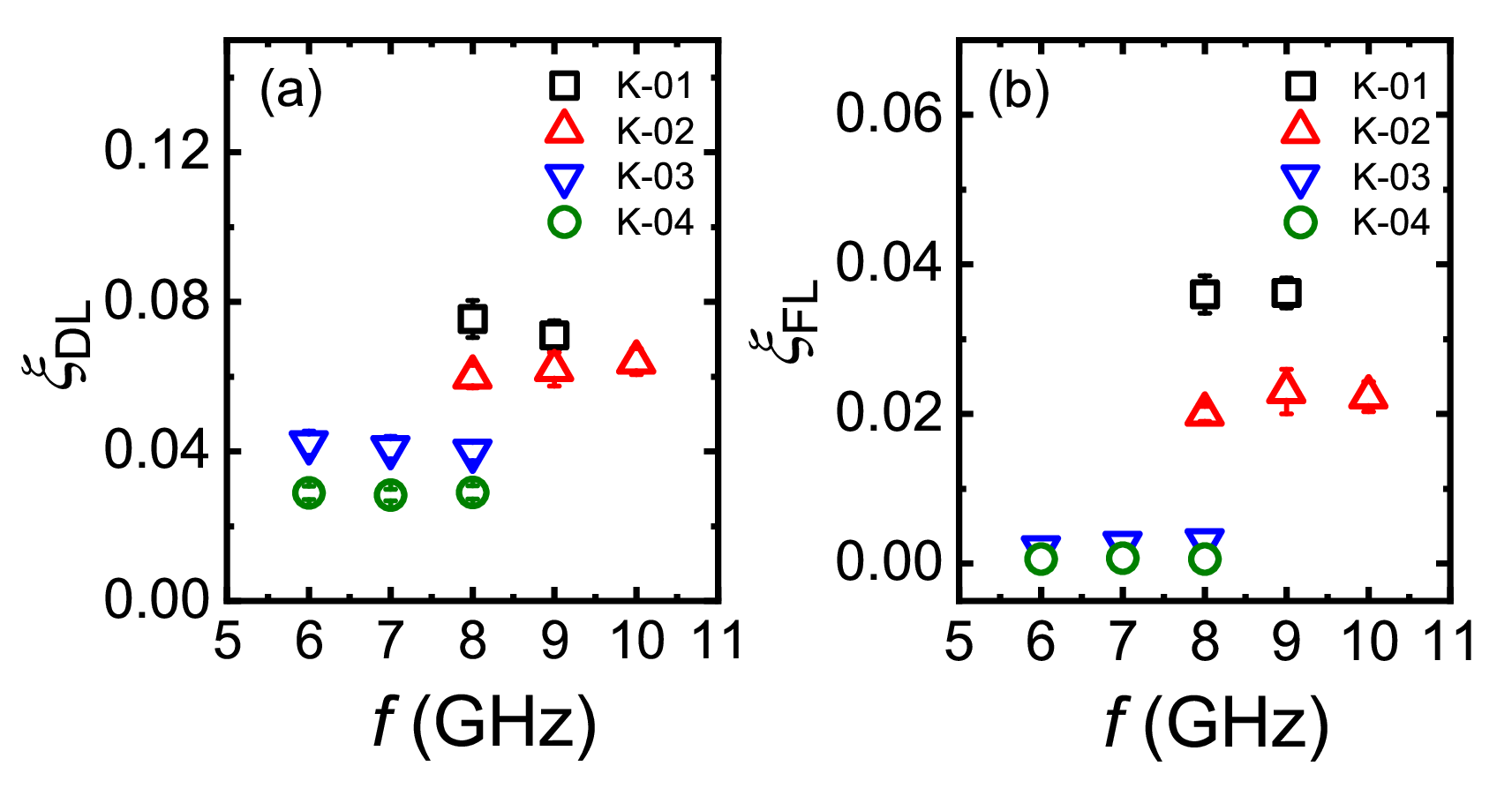}}
\caption{\label{fig:ch3_eff}Efficiency for damping like ($\xi_{DL}$) and field like ($\xi_{FL}$) torques per applied current density $j$ calculated using Eq.~\ref{eq:efficiency} at different frequencies.}
\end{figure}

Using the value of $H_\textrm{DL}$( $H_{\textrm{FL+Oe}}$) we estimated the DL (FL) efficiency per unit applied current density as:
\begin{align}
    \xi_\textrm{DL(FL+Oe)} = \left(\frac{2e}{\hbar}\right)\frac{\mu_0M_{\textrm s}d_{\textrm{FM}}H_{\textrm{DL(FL+Oe)}}}{j}
    \label{eq:efficiency}
\end{align}
\begin{table}
\centering
\small
 \caption{\label{tab:sot2} Efficiency magnitude for damping like ($\xi_{\textrm{DL}}$) and field like ($\xi_{\textrm{FL}}$) torques per applied current density $j$, spin Hall conductivity ($\sigma_{SH})$ and the saturation magnetisation ($\mu_0M_{\textrm{S}}$) of the samples.}
\begin{tabular}{ccccc}
\hline\hline
Parameter &  &sample  &  & \\
 &K-01 &K-02 &K-03 &K-04\\
\hline
$\xi_{\textrm{DL}}$ &0.08  &0.06  & 0.04  & 0.03 \\
$\xi_{\textrm{FL}}$ &0.04  &0.02  & 0.003  & 0.0004 \\
$\sigma_{SH}((\hbar/2e)10^3 \Omega^{-1} cm^{-1})$ &7.0 &4.6 & 2.2 & 1.0\\
$\mu_0 M_{\textrm{s}} (T)$ &0.362 &0.331 &0.377 &0.298\\
\hline
\hline
\end{tabular}
\end{table} 
where $d_\textrm{FM}$ and $j$ are thickness of Ni and the applied current density respectively. The values of $\xi_{\textrm{DL}}$ and $\xi_{\textrm{FL}}$ are summarised in Table~\ref{tab:sot2}. It is found that the values of $\xi_{\textrm{DL}}$ and $\xi_{\textrm{FL}}$ are largest for the multi-layer and epitaxial films. For both multi-layer and bi-layer films, the SOT efficiency is larger when grown epitaxially on the sapphire substrate. Experimental results at different frequencies (see Figs.~\ref{fig:ch3_eff} and \ref{fig:ch3_sot2}) support reproducibility of $\xi_{\textrm{DL}}$ and $\xi_{\textrm{FL}}$.

\begin{figure}
\centering
{\includegraphics[width=0.48\textwidth]{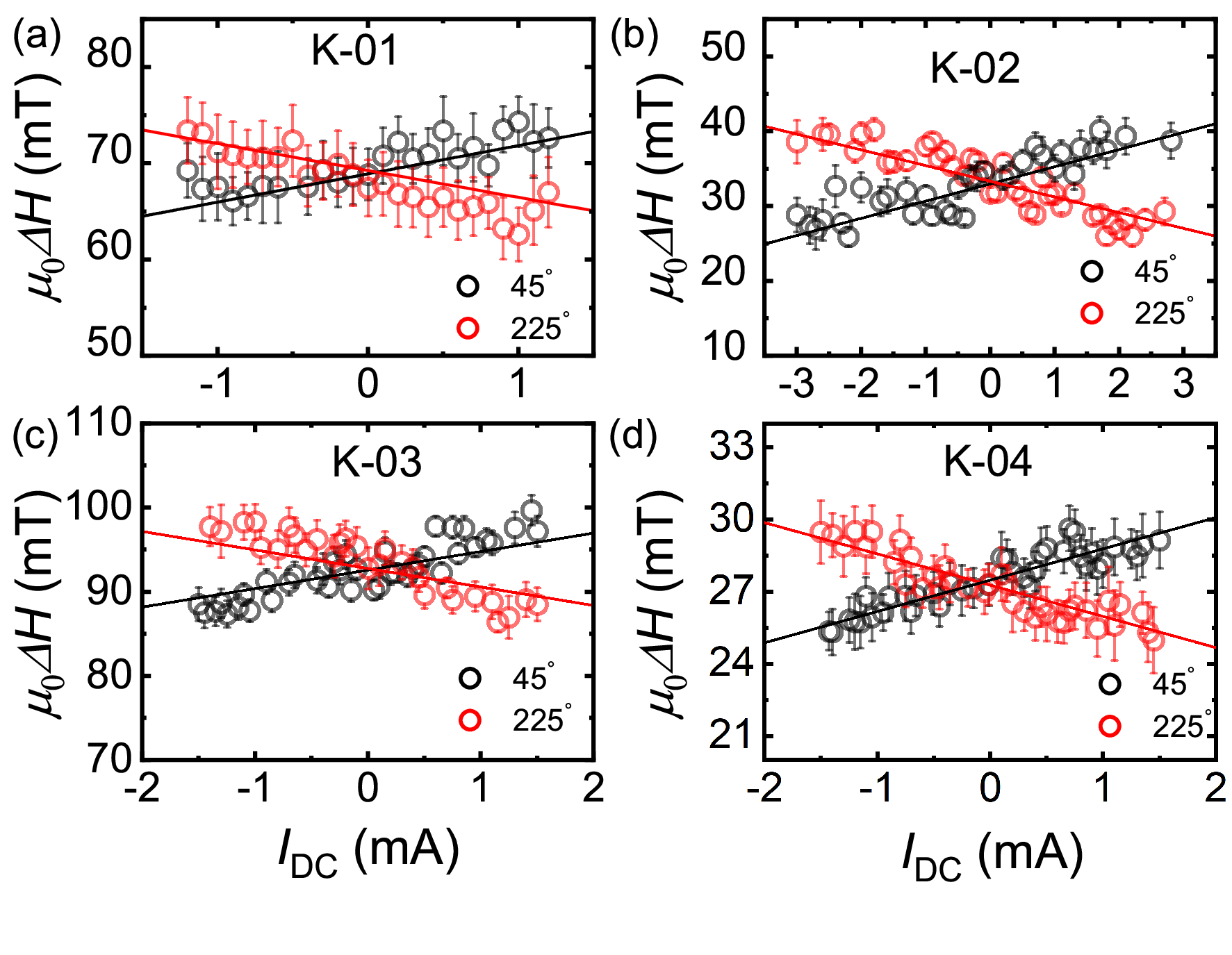}}
\caption{\label{fig:ch3_dh}The change of the FMR linewidth as a function of injected current $I_\textrm{DC}$ at frequency of 8 GHz for K-01, K-02 and 6 GHz for K-03, K-04.}
\end{figure}

\begin{figure}
\centering
{\includegraphics[width=0.48\textwidth]{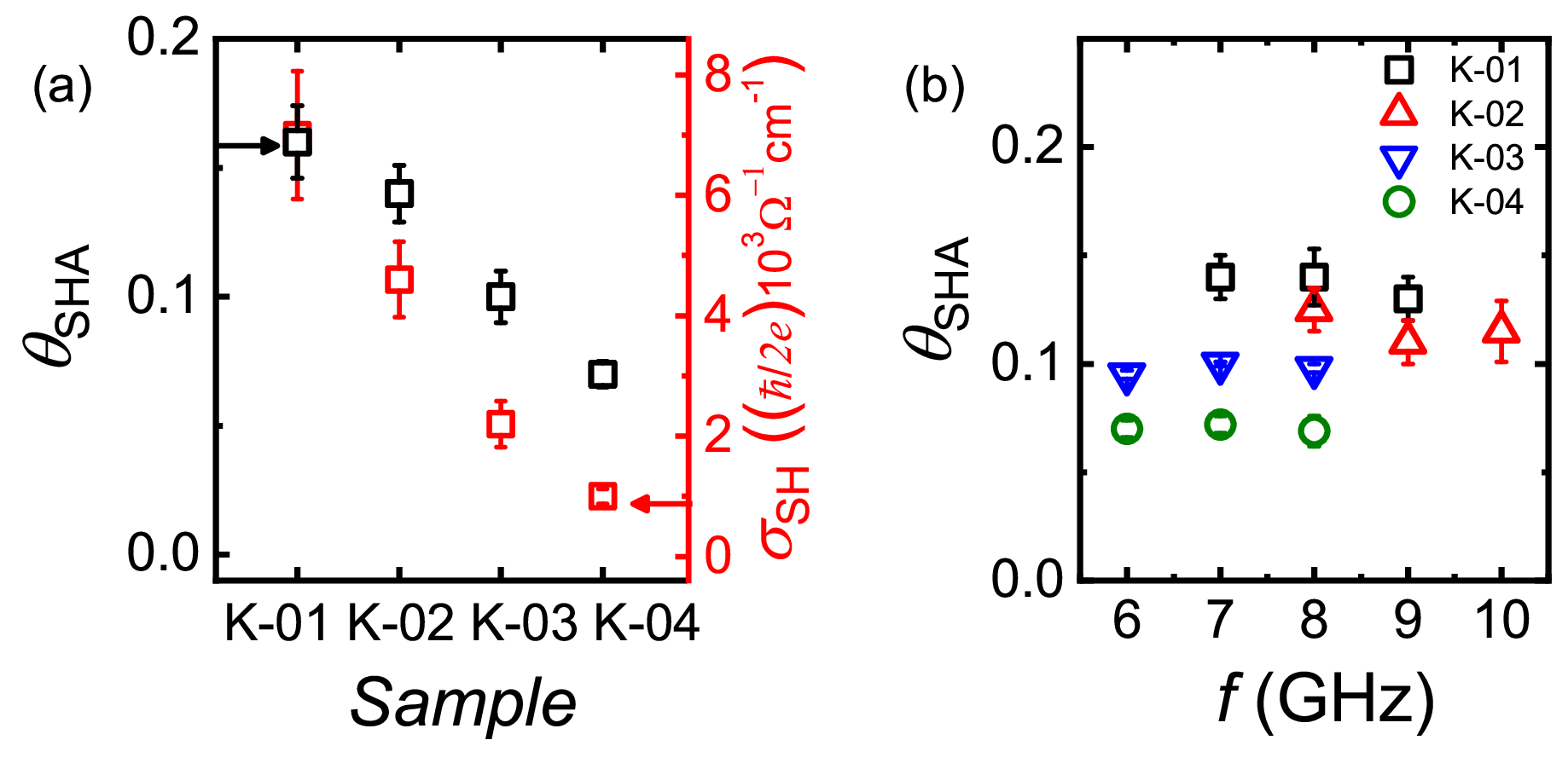}}
\caption{\label{fig:ch3_sha} $\theta_\textrm{SHA}$ calculated from (a) Slope of Fig. 7 for all samples (b) from slope of power dependence (Fig. 13) at different frequencies. The results calculated for bi-layer samples are consistent for all three methods. It can be seen that epitaxial samples have larger $\theta_\textrm{SHA}$ compared to poly-crystalline samples thus indicating the role of crystal structure in modifying SOT. The right axes of (a) shows the spin Hall conductivity $\sigma_{SH}$ calculated using the value of $\theta_\textrm{SHA}$.}
\end{figure}

To characterize the SOT efficiency, the FMR linewidth $\mu_0\Delta H$ was measured as a function of dc current, similar to technique used in \cite{liu2011spin} and the results are shown in Fig.~\ref{fig:ch3_dh}.
A clear linewidth broadening is observed for both  $\theta = 45\degree$ and $\theta = 225\degree$. The macrospin model can provide the linewidth broadening as \cite{liu2011spin}: 
{\small
\begin{align}
\mu_0\Delta H&=\frac{2 \pi f}{\gamma}\left(\alpha+\frac{\sin \theta}{\left(H_{\mathrm{ext}}+0.5 M_{\mathrm{eff}}\right)\mu_{0}  M_{\textrm{S}} d_{\textrm{FM}}} \frac{\hbar \theta_{\textrm{SHA}} J_{\textrm{C}}}{2 e}\right)
\label{eq:exp12}
\end{align}}
In Eq.~\ref{eq:exp12} the terms $\gamma$, $e$, $\hbar$, $d_\textrm{FM}$ and $H_\textrm{ext}$  are gyromagnetic ratio, electronic charge, reduced Planck's constant, thickness of ferromagnet and magnetic field, respectively. Using Eq.~\ref{eq:exp12} we calculated the effective spin Hall angle given as $\theta_\textrm{SHA} = J_{\textrm s}/J_{\textrm c}$ where $J_{\textrm s}$ is the spin current density in the Pt layer and $J_{\textrm c}$ is the charge current density. $\theta_\textrm{SHA}$ takes into account not only the contribution from bulk spin Hall effect but also other contributions originating from spin-orbit coupling as discussed in section IV. $\theta_\textrm{SHA}$ for each sample is presented in Fig.~\ref{fig:ch3_sha} (a). We can see similar trend of the values of $\theta_\textrm{SHA}$ and $\xi_{\textrm{DL}}$ with respect to the sample supporting our claim that multi-layer and epitaxial growth enhances the SOTs.    

In the multi-layer samples, we still observe sizable values of $\theta_\textrm{SHA}$. We discuss the details about this enhancement in multi-layer samples in discussion section. We therefore use Eq.~\ref{eq:exp12} to characterise averaged torques generated in the entire FM layers, from which the efffective spin-Hall angle in the entire device can be still discussed. We would like to mention here that the $\theta_\textrm{SHA}$ for multi-layer epitaxial samples $\sim$ 0.15 is larger than previous reported values \cite{feng2012spin,kondou2012evaluation,obstbaum2014inverse}. This provides a new route for enhancement of $\theta_\textrm{SHA}$ through growth. 



To validate the conclusions drawn we used another method for obtaining $\theta_\textrm{SHA}$. This was done by calculating the slope of the power dependence of $V_\textrm{sym}$ (see Appendix E and Fig.~\ref{fig:power} ) using the method given in \cite{sud2021parity}. We show the calculated values of $\theta_\textrm{SHA}$ in Fig.~\ref{fig:ch3_sha} (b). We obtained the values similar to that obtained by linewidth modulation with current. Thus, we can confirm that  multi-layer epitaxial samples show large $\theta_\textrm{SHA}$. Using the value of $\theta_\textrm{SHA}$ we calculated the spin Hall conductivity\cite{conductivity} and summarise the values in Table.~\ref{tab:sot2} and Fig.~\ref{fig:ch3_sha}(a). It is seen that the largest spin Hall conductivity is exhibited by epitaxial multi-layer samples which also show the largest spin Hall angle. Based on the above results we can claim that crystal orientation effects and growth play an important role in efficient SOT generation. 
\subsection{Spin-Orbit Torque harmonic Hall measurements}
We also calculated the SOT efficiency by measuring harmonic Hall resistance. 
\begin{figure}
\centering
{\includegraphics[width=0.47\textwidth]{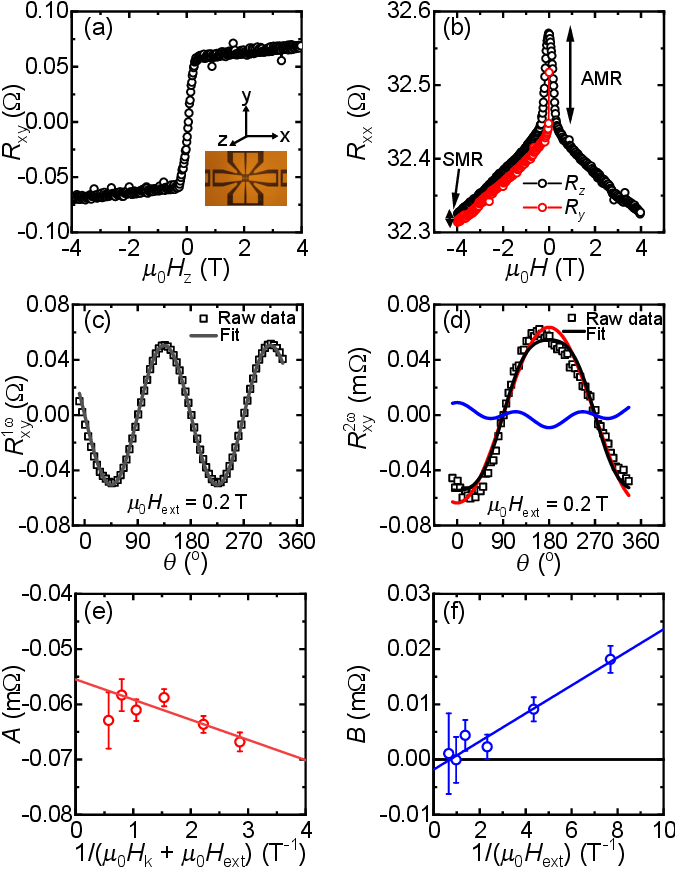}}
\caption{\label{fig:ch3_harmonic}(a) Transverse Hall resistance, $R_{\textrm{xy}}$, for the Hall-bar device measured with field applied along the out-of-plane direction. (b) Longitudinal resistance, $R_{\textrm{xx}}$, for the same device measured for the field applied along in-plane (red) and out-of-plane (black) directions. Angular dependence of  harmonic Hall resistance at external fields of  0.2 T for (c)  $R^{1\omega}_{\textrm{xy}}$  and (d) $R^{2\omega}_{\textrm{xy}}$. Black lines are the fit to the experimental data  using $\sin\theta$ and Eq.~\ref{eq:3} for (c) and (d) respectively. In (d) red and blue lines correspond to the $\cos\theta$ component (A) and $\cos 2\theta\cos\theta$ component (B) respectively. (e) Prefactor A of the $\cos\theta$ component and (f) prefactor B of the $\cos 2\theta\cos\theta$ component as a function of 1/($H_{\textrm k}+H_{\textrm {ext}}$) and 1/($H_{\textrm {ext}}$), respectively. Solid lines are the linear fit to the data from which the values of $H_\textrm{DL}$ and $H_{\textrm{FL+Oe}}$ are obtained. }
\end{figure}
Inset of Fig.~\ref{fig:ch3_harmonic} (a) shows the device pattern for harmonic-Hall measurements fabricated on sample K-01. The average resistivity was found to be 26 $\mu\Omega$ cm. The measurements were performed at a frequency of 172 Hz and a sinusoidal current excitation, $I_{\textrm{o}}\sin(\omega t)$ with amplitude of $I_{\textrm{o}}$ = 7  mA was applied using an alternating current source meter. Figure~\ref{fig:ch3_harmonic} shows (a) transverse Hall and (b) longitudinal resistance measured as a function of applied external field. We can see from Fig.~\ref{fig:ch3_harmonic} (b) that the Hall resistance is dominated by planar Hall effect of the order of 0.4\% as compared to very small spin Hall magnetoresistance (0.05 \%). We estimated the value of perpendicular anisotropy field, $\mu_0 H_{\textrm{k}}$ = 0.25 T, by linear fitting of the experimental data shown in Fig.~\ref{fig:ch3_harmonic}(a) at high fields. We were able to quantify the effective DL-SOT and FL-SOT of the Ni film by second harmonic-Hall resistance ($R_{2\omega}$) measurements. The sample was rotated to vary in-plane field orientations as shown in Fig.~\ref{fig:ch3_harmonic} (d)  for the applied external field of 0.2 T. The data points are fitted using the standard equation for the second harmonic-Hall resistance given by \cite{seki2021spin}:
\begin{align}
R_{2\omega} &= \frac{1}{2}\left(R_{\textrm{AHE}}\frac{H_{\textrm{DL}}}{H_{\textrm{k}}+H_{\textrm{ext}}}+R_{\textrm{const}}\right)\cos\theta\nonumber\\
&-\left(R_{\textrm{PHE}}\frac{H_{\textrm{FL+Oe}}}{H_{\textrm{ext}}}\right)\cos 2\theta\cos \theta\nonumber \\
&= A\cos\theta + B\cos 2\theta\cos\theta ,
\label{eq:3}
\end{align}
where, $\theta$ is the angle between applied external field and current. $R_\textrm{AHE}$, $R_\textrm{PHE}$, $H_{\textrm k}$, $H_{\textrm {DL}}$ and $H_{\textrm {FL+Oe}}$ are the anomalous Hall resistance, planar Hall resistance, perpendicular anisotropy field, DL spin-orbit effective field and combined FL spin-orbit and Oersted effective field respectively. $R_\textrm{const}$ is the component of the $R_{2\omega}$  which is independent of the applied field. The fitted curves in Fig.~\ref{fig:ch3_harmonic} (d) show the contribution from the $\cos\theta$ component (prefactor A: red line) and $\cos 2\theta\cos\theta$ component (prefactor B: blue line). The resultant fit (black line) is the sum of two components. The values of prefactors A and B obtained from fittings are plotted as a function of 1/($H_{\textrm k}+H_{\textrm{ext}}$) and 1/($H_{\textrm{ext}}$) in  Fig.~\ref{fig:ch3_harmonic} (e) and (f), respectively. It can be seen that as the field is increased, the FL contribution diminishes and only the DL contribution remains. The linear fits to the plots give the value of $H_\textrm{DL}$ and $H_{\textrm{FL+Oe}}$ from which we estimate the DL ( FL) efficiency using Eq.~\ref{eq:efficiency}. The value of $\xi_{\textrm{DL}} \approx 0.08 \pm 0.01$. The  FL contribution, $\xi_{\textrm{FL}} \approx 0.03 \pm 0.003$ which can be explained as entirely coming from Oersted field due to 1 nm thick Pt layer. The DL and FL efficiency are similar to value calculated from FMR measurements. The consistency of the DL and FL efficiency values obtained from both methods justify the claim that multi-layer epitaxial samples show better DL and FL efficiency. A detailed discussion and possible reasoning for variation of efficiency with multi-layer growth and crystallinity follow in the next section.


\section{Discussion}
In this section we discuss possible sources of large DL and FL torque generation efficiencies in multi-layer and epitaxial samples. Both epitaxial samples (K-01 and K-03) show larger $\xi_{\textrm{DL}}$ than the polycrystalline counterparts. While a number of parameters contribute to $\xi_{\textrm{DL}}$, we speculate a few as follows. The momentum scattering rate is expected to be different between polycrystalline and epitaxial samples due to different numbers of e.g. defects, crystallographic domain boundaries and chemical disorders. This difference naturally leads to the variation of resistivity, the extrinsic spin-Hall effect as well as the spin-relaxation rate. Nguyen et al.\cite{nguyen2016spin} and Lee et al.\cite{Lee_PRB2017} systematically demonstrated that the spin-transport parameters significantly vary with Pt resistivity. In addition, we would like to mention that epitaxial and polycrystalline Pt films display dissimilar relationship between the spin and momentum scattering rates, attributed to EY and DP spin-relaxation mechanisms \cite{ryu2016observation}. This difference might play a role for $\xi_{\textrm{DL}}$ since $\xi_{\textrm{DL}}$ is a device parameter including the spin-relaxation rate in our model. Furthermore, electrons in epitaxial films flow along the specific crystallographic orientation in our devices. The spin-orbit property of electrons populated around the corresponding momentum point in the Fermi surface determines $\xi_{\textrm{DL}}$. This is not the case for the polycrystalline samples where $\xi_{\textrm{DL}}$ has contributions from all momentum points across the entire Fermi surface. Indeed, the notable difference in spin-torque generation between epitaxial and polycrystalline Co/Pt samples has been reported by Ryu et al.\cite{ryu2019crystal}.  

In our experiments, the magnitude of DL torque efficiency is always larger in multi-layer samples than bi-layers. This is unexpected because both DL and FL torques should be cancelled in a Ni layer sandwiched by two Pt layers, i.e. those in our multi-layers where only the bottom-most Ni layer is expected to be torqued when we apply a current. This does not seem to be the case in our devices where sizable spin torques were experimentally observed, suggesting that the spin-torque properties in our multi-layer device are not as simple as the ideal case. For example, the quality of top and bottom interfaces cannot be necessarily the same due to different magnitudes of intermixing and/or strain propagation at the interfaces \cite{kim2017role, kim2016asymmetric}. This is supported by our X-ray reflectivity (XRR) results presented in Appendix A, where we are required to introduce intermixing (NiPt alloy) layers in order to fit the experimental data well for epitaxial multi-layer samples, whereas poly-crystalline samples do not need such addition for better fitting. The best-fit parameters indicate that the upper and lower interfaces sandwiching the ferromagnetic layer are not equivalent in multi-layer samples. The different interface quality hence suggests non-vanishing spin currents injected from the top and bottom Pt layers, exerting measurable torques in our experiments. It has been shown experimentally that DL and FL torques are modified by consequent intermixing \cite{garello2013symmetry} and insertion of spacer layers \cite{fan2013observation} between the ferromagnet and the non-magnet. The upper/lower interface quality would also lead to asymmetric interfacial spin absorption that potentially contributes to the large SOT magnitude in multi-layer samples. For multi-layer growth, the film quality tends to be improved with growth due to lesser contributions of lattice mismatch and the roughness from the substrate as growth continues. The resultant sharp interfaces may lead to enhancement of SOTs as shown in previous works \cite{choi2022thickness1,choi2022thickness2}. 

 Finally we consider other current-induced spin-torque generation mechanisms\cite{amin2016spin,amin2016spin2,kurebayashi2014antidamping}. It is possible to imagine self-induced torques \cite{wang2019anomalous,soya2023isotropic1} due to spin current flowing in the Ni layers, where the asymmetry in top and bottom interfaces can induce an additional SOT\cite{wang2019anomalous}. We cannot rule out the possibility of potential contributions from orbital Hall effect due to the long range orbital current generated in Pt\cite{go2023long, hayashi2023observation}.\\

\section{Conclusion}
In this study we presented the measurement of the DL and FL torque efficiencies in  Ni/Pt multi-layer and bi-layer samples using techniques of SOT-FMR and harmonic Hall resistance. The effects of layer stacking and crystal orientation on SOT was examined. We found that the DL torque efficiency ($\xi_\textrm{DL}$) is enhanced by multi-layer epitaxial growth and the value of $\xi_\textrm{DL}$ per unit applied current density obtained for epitaxial multi-layer samples ( $0.08 \pm 0.01$) $\sim$ 160\% larger than the value for bi-layer poly-crystalline samples ( $0.03 \pm  0.002$). The results show a large spin Hall angle in multi-layer samples, without cancellation coming from the symmetry of stacking structure, which is attributable to the different interface qualities between upper and lower interfaces. The value of $\theta_\textrm{SHA} \sim 0.15$ in multi-layer epitaxial films is larger than previous reported values. These findings provide a new route for enhancement of DL SOT efficiency through multi-layer epitaxial growth. The results also indicate that epitaxial samples show larger $\xi_\textrm{FL}$ compared to the poly-crystalline samples. Apart from the bulk spin Hall and interfacial effects, some other sources such as self-induced SOTs in Ni and orbital Hall effect can potentially play an important role in SOT generation. Although we could not experimentally separate the different contributions to SOTs, the relatively large efficiency of spin current generation that is observed is promising for applications that utilize SHE to manipulate ferromagnetic dynamics. Our results shed light on the importance of crystal orientation, multi-layering and interface quality for enhancing the magnitude of SOT which is important for spintronic devices. 

\section*{Acknowledgements}
 A. S. thanks  JSPS Postdoctoral fellowship for research in Japan (P21777) and EPSRC for their supports through NPIF EPSRC Doctoral studentship (EP/R512400/1). T. S. is supported by Grant-in-Aid for Scientific Research (A) (JP20H00299) and Grant-in-Aid for Scientific Research (S) (JP18H05246)  from JSPS KAKENHI. 
\appendix
\begin{figure}
\centering
{\includegraphics[width=0.45\textwidth]{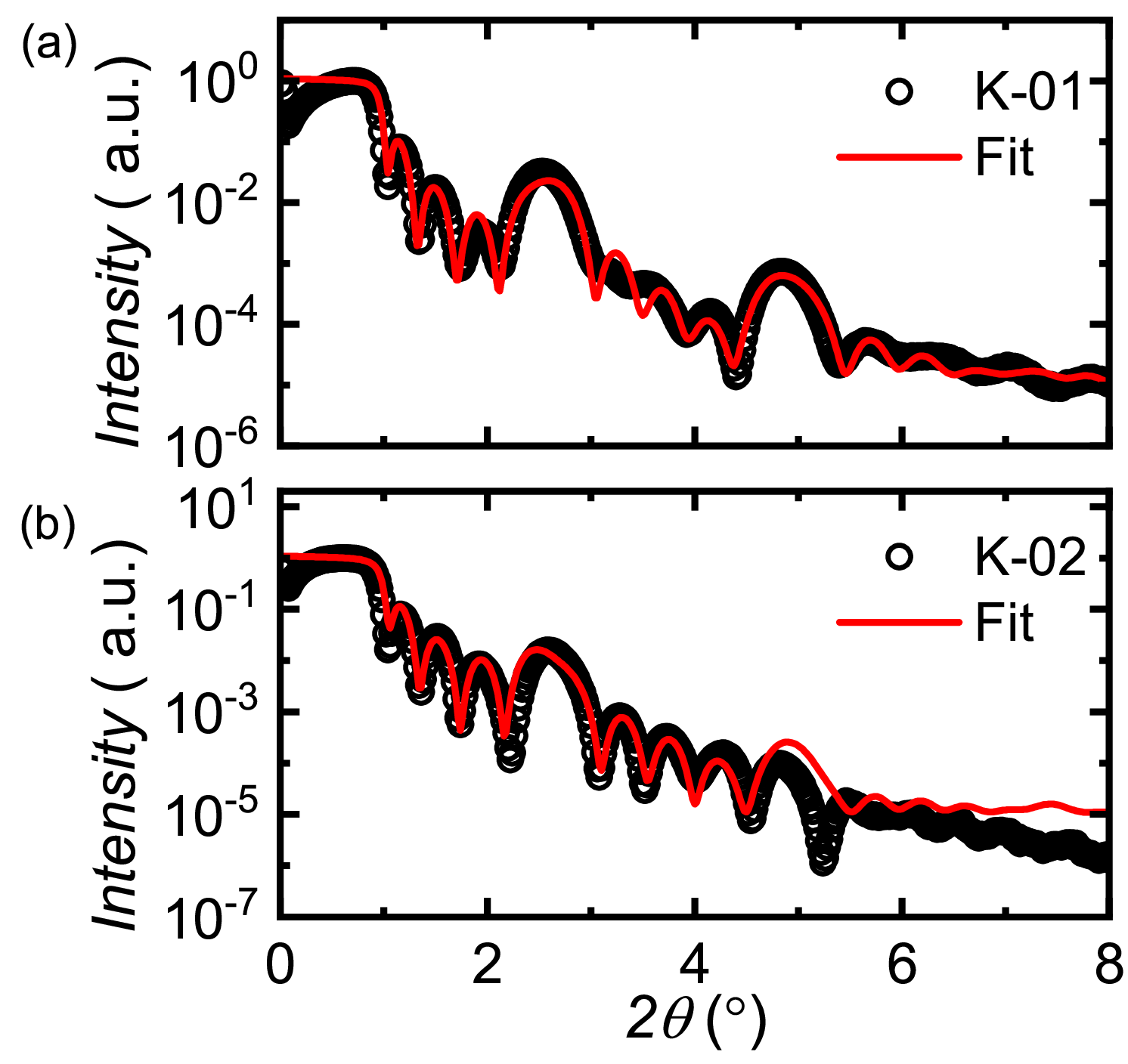}}
\caption{\label{fig:xrr} X-ray reflectivity profiles for multi-layer samples. Solid lines are fitting results. }
\end{figure}

\section{XRR measurements}
In order to determine the thickness of the films and to determine the surface roughness, XRR measurements were performed as shown in Fig.~\ref{fig:xrr} using scintillation counter. The good quality of films is confirmed from the presence of large number of fringes. The data points were fitted by the Globalfit software of Rigaku using a multi-layer structure consisting of Ni and Pt layers. The fitting parameters are summarised in Table.~\ref{tab:appendix}. The surface roughness obtained was $\sim$ 0.5 nm. For the  sample K-01, existence of intermixing layer (i.e. a very thin Ni-Pt alloy layer) is assumed in order to fit the experimental XRR. Without considering the intermixing layer, the experimental XRR cannot be fitted numerically. This intermixing is attributable to the substrate deposition temperature of 400$\degree$C. This might be related to the higher DL torque for the epitaxial Ni/Pt. For polycrystalline K-02 sample, an intermixing layer (i.e. Ni-Pt alloy layer) is not taken into account because the room temperature deposition for K-02 did not promote intermixing. Even without intermixing layer, the experimental XRR is fairly fitted numerically.

\begin{table}
\centering
 \caption{\label{tab:appendix} Fitting parameters for X-ray reflectivity measurements }
\begin{tabular}{ccccc}
\hline\hline
Sample &Layer &Density &$d (nm)$ & $\sigma (nm)$ \\
 & &(g/cm$^3$) &thickness &roughness \\
\hline
&Pt &21.5$\pm$0.01 &1.0 $\pm$0.03 &1.24 $\pm$0.03 \\
&NiPt &11.9$\pm$0.06 & 0.4 $\pm$0.02&0.17 $\pm$0.01 \\
$K-01$&Ni &8.9$\pm$0.01 &1.9 $\pm$0.01 &0.11 $\pm$0.02\\
&NiPt &11.5$\pm$0.04 &0.6 $\pm$ 0.02 &0.09 $\pm$0.01 \\
&Al$_2$O$_3$ &1.8$\pm$0.01 &3.1$\pm$0.01 &0.18$\pm$0.01 \\
\hline
&Pt &21.5$\pm$0.01&1.1 $\pm$0.02 &1.3 $\pm$0.02 \\
$K-02$&Ni &8.9$\pm$0.04 &2.5 $\pm$0.02 &0.4 $\pm$0.01\\
&Al$_2$O$_3$ &1.6$\pm$0.03 &2.8$\pm$0.05&0.1$\pm$0.01 \\
\hline\hline
\end{tabular}
\end{table}

\section{XRD profile calculations}
In the case of metallic superlattice, the step model is applicable to explain the peak positions for the experimental XRD profile. The x-ray scattering intensity $(I(Q))$ for the metallic superlattice consisting of $\mathrm{Ni}$ and $\mathrm{Pt}$ is given by $I(Q)=I_e\left|F_{\mathrm{Ni}}(Q)+F_{\mathrm{Pt}}(Q) \exp \left(i Q D_{\mathrm{Ni}}\right)\right|^2\left|\sum_{k=0}^{N-1} \exp (i Q k \Lambda)\right|^2$ where $I_e$ is Thomson scattering intensity, $F_{\mathrm{Ni}(\mathrm{Pt})}(Q)$ is the structural factor of $\mathrm{Ni}(\mathrm{Pt})$, $D_{\mathrm{Ni}}$ is the thickness of the Ni layer, $\Lambda$ is the superlattice period, and $Q$ is the scattering vector. $\left|\sum_{k=0}^{N-1} \exp (i Q k \Lambda)\right|^2$ corresponds to Laue function $L(Q)$, and is expressed as $L(Q)=\left|\sum_{k=0}^{N-1} \exp (i Q k \Lambda)\right|^2=\frac{\sin ^2\left(\frac{N Q \Lambda}{2}\right)}{\sin ^2\left(\frac{Q \Lambda}{2}\right)}$ where $N$ is the repetition number. The term of structural factor is expressed as

{\small
$\begin{aligned} 
&\mid F_{\mathrm{Ni}}(Q) +\left.F_{\mathrm{Pt}}(Q) \exp \left(i Q D_A\right)\right|^2=\left|F_{\mathrm{Ni}}(Q)\right|^2+\left|F_{\mathrm{Pt}}(Q)\right|^2\\&+F_{\mathrm{Ni}}(Q) F_{\mathrm{Pt}}{ }^*(Q) \exp \left(-i Q D_{\mathrm{Ni}}\right)\\&+F_{\mathrm{Ni}}{ }^*(Q) F_{\mathrm{Pt}}(Q) \exp \left(i Q D_{\mathrm{Ni}}\right).
\end{aligned}$\\
}
Using the areal atomic density $\left(\eta_{\mathrm{Ni}}\right.$ and $\left.\eta_{\mathrm{Pt}}\right)$, lattice spacing $\left(d_{\mathrm{Ni}}\right.$ and $\left.d_{\mathrm{Pt}}\right)$, atomic scattering factor$\left(f_{\mathrm{Ni}}\right.$ and $\left.f_{\mathrm{Pt}}\right)$ and number of lattice plane $\left(n_{\mathrm{Ni}}\right.$ and $n_{\mathrm{Pt}}$ ), the term of structural factor can be transformed into

$$
{\footnotesize
\begin{aligned}
&|F(Q)|^2=f_{\mathrm{Ni}}^2(Q) \eta_{\mathrm{Ni}}^2 \frac{\sin ^2\left(\frac{n_{\mathrm{Ni}} Q d_{\mathrm{Ni}}}{2}\right)}{\sin ^2\left(\frac{Q d_{\mathrm{Ni}}}{2}\right)}+f_{\mathrm{Pt}}^2(Q) \eta_{\mathrm{Pt}}^2 \frac{\sin ^2\left(\frac{n_{\mathrm{Pt}} Q d_{\mathrm{Pt}}}{2}\right)}{\sin ^2\left(\frac{Q d_{\mathrm{Pt}}}{2}\right)} \\
&+2 f_{\mathrm{Ni}}(Q) f_{\mathrm{Pt}}(Q) \eta_{\mathrm{Ni}} \eta_{\mathrm{Pt}} \frac{\sin \left(\frac{n_{\mathrm{Ni}} Q d_{\mathrm{Ni}}}{2}\right)}{\sin \left(\frac{Q d_{\mathrm{Ni}}}{2}\right)} \frac{\sin \left(\frac{n_{\mathrm{Pt}} Q d_{\mathrm{Pt}}}{2}\right)}{\sin \left(\frac{Q d_{\mathrm{Pt}}}{2}\right)} \cos \left(\frac{\Lambda Q}{2}\right)
\end{aligned}}%
$$

Then, the $\mathrm{x}$-ray scattering intensity was calculated for $[\mathrm{Ni}(3 \mathrm{~nm}) / \mathrm{Pt}(1 \mathrm{~nm})]_{\mathrm{x} 5}$ with the (111) crystal orientation.

\section{Microwave Calibration}
\begin{figure*}
\centering
{\includegraphics[width=0.8\textwidth]{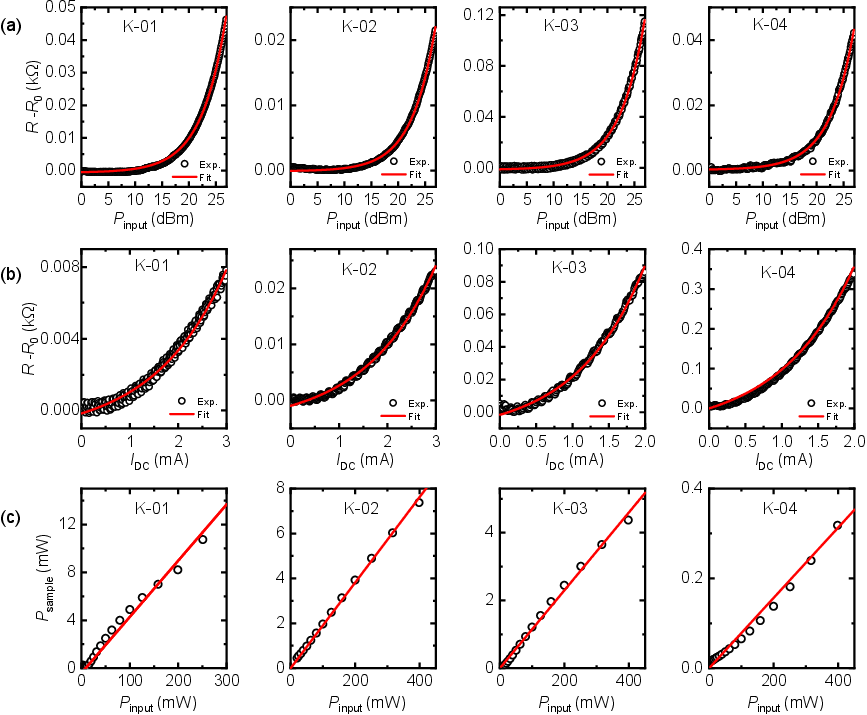}}
\caption{\label{fig:ch3_heat} (a) Resistance change as a function of microwave power for four set of samples. (b) Resistance change as a function of dc current and (c) Power from source as a function of the power in the sample for different samples. }
\end{figure*}

 There is a large impedance mismatch between microwave lines and the sample (with few thousand $\Omega$ in resistance) which causes a large amount of power reflection from the devices. As a result the amount of power reaching the sample is a fraction of power supplied from the source. To quantify the actual power reaching the device we used a bolometric technique \cite{fang2011spin,kurebayashi2014antidamping} in which we compared the resistance change caused due to joule heating when a known
dc current, $I_{\textrm{DC}}$  flows in the sample with that caused by flow of microwave power, $P_{\textrm{input}}$. Figs.~\ref{fig:ch3_heat} shows the resistance change by two excitations. The current flowing through the device ($I_\textrm{DC}$) is quantified at given microwave frequency and then the microwave power at sample is calculated by scaling it with sample resistance. Figs.~\ref{fig:ch3_heat}(c) shows the microwave power at sample  plotted against input microwave power ($P_\textrm{input}$). It can be seen that that power input at the sample is fraction of the power from the microwave source. For samples K-03 and K-04 which had higher resistance, the current reaching the sample is smaller as compared to K-01 and K-02.

\section{Fitting Equation for FMR Data}
The expression for DC signal from spin rectification used for fitting the FMR data is given as:
\begin{align}
\mathrm{V}_{\mathrm{dc}}&=V_{\mathrm{sym}} \frac{\Delta H^{2}}{\left(H_{\mathrm{ext}}-H_{\mathrm{res}}\right)^{2}+\Delta H^{2}}\nonumber\\
&+ V_{\mathrm{asym}} \frac{\left(H_{\mathrm{ext}}-H_{\mathrm{res}}\right) \Delta H}{\left(H_{\mathrm{ext}}-H_{\mathrm{res}}\right)^{2}+\Delta H^{2}}
\label{eq:voltage}
\end{align}
where \(V_{\text {sym}}\) and \(V_{\text {asym}}\) are the Lorentzian components in symmetric and anti-symmetric lineshape given below. 
\begin{align}
V_{\mathrm{sym}}&=\frac{\mathrm{I}_{0} \Delta \mathrm{R}}{2} \frac{\omega}{\mu_{0} \gamma \Delta H\left(2 H_{\mathrm{res}}+H_{1}+H_{2}\right)} h_{z} \sin 2 \theta\nonumber\\&=\frac{\mathrm{I}_{0} \Delta \mathrm{R}}{2} \mathrm{~A}_{\mathrm{sym}} h_{z} \sin 2 \theta 
\label{eq:F13}
\end{align}
\begin{align}
V_{\mathrm{asym}}&=\frac{\mathrm{I}_{0} \Delta \mathrm{R}}{2} \frac{\left(H_{\mathrm{res}}+H_{1}\right)}{\Delta H\left(2 H_{\mathrm{res}}+H_{1}+H_{2}\right)}(-h_{\mathrm{x}} \sin \theta\nonumber\\
&+ h_{\mathrm{y}} \cos \theta) \sin 2 \theta\notag\\
&=\frac{\mathrm{I}_{0} \Delta \mathrm{R}}{2} \mathrm{~A}_{\mathrm{asy}}\left(-h_{\mathrm{x}} \sin \theta+h_{\mathrm{y}} \cos \theta\right)
\label{eq:F14}
\end{align}
 The terms $A_{\textrm{sym}}$ and $A_{\textrm{asy}}$ are the scalar amplitudes of the magnetic susceptibility $\left(A_{\textrm i} = \chi _{\textrm i}/M_{\textrm s}\right)$ and depend on the magnetic anisotropy of the device. The other terms are the external field $(H_{\rm{ext}})$ and $H_1, H_2$ are the terms containing demagnetisation field and in-plane/out-of-plane anisotropy fields. $h_{\mathrm{x}}, h_{\mathrm{y}}, h_{\mathrm{z}}$ are the components of current-induced effective field at microwave frequency, $f$, which drives the magnetic moments given as: \(\mathbf{h}_{\text {eff }}=\left(h_{\mathrm{x}}, h_{\mathrm{y}}, h_{\mathrm{z}}\right) \mathrm{e}^{j 2\pi f t}\). \(\gamma\), \(M_{\mathrm{s}}\), $H_\textrm{res}$, $\Delta H$, $I_0$ and $\Delta R$ are the gyromagnetic ratio, saturation magnetisation, resonance field, half width at half maximum linewidth of resonance, current amplitude in the device and AMR resistance change respectively. 

\begin{figure}
\centering
{\includegraphics[width=0.48\textwidth]{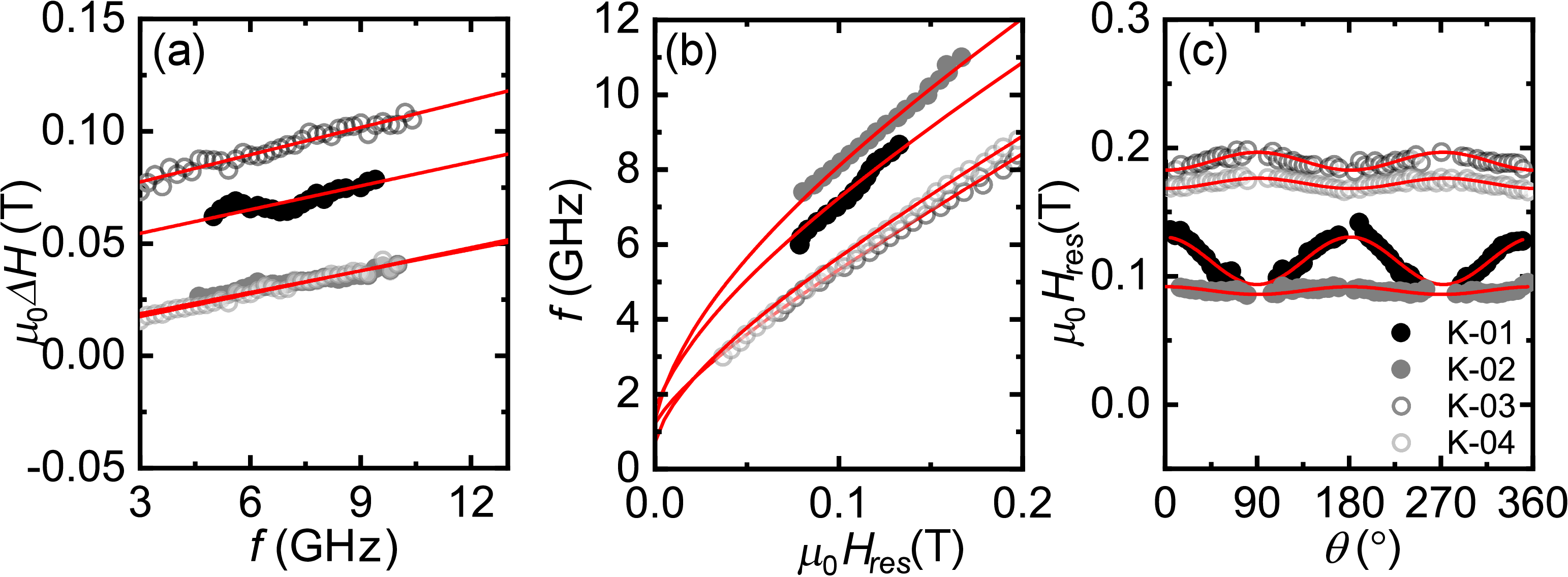}}
\caption{\label{fig:freq}(a) Frequency dependence of the half-width at half maximum (HWHM) linewidth at $\theta =45 \degree$ for all the samples. The inhomogeneous broadening is obtained from the intercept by fitting the data with linear-fit and the slope gives the Gilbert damping component, $\alpha$. (b) Resonance field, $\mu_0H_\textrm{res}$ as a function of frequency at an angle $\theta = 45\degree$ for all samples. The solid lines are the fittings. (c) In plane angular dependence of resonance field, $\mu_0H_{\textrm{res}}$ obtained from fitting the FMR scans for different set of samples measured at 8 GHz.}
\end{figure}

The anisotropy fields used for obtaining $H_1, H_2$ are calculated by fitting the frequency dependence of resonance field shown in Fig.~\ref{fig:freq} (b) using  Kittel resonance formula \cite{PhysRev.73.155}. The in-plane uniaxial($\mu_0H_{2\parallel}$) and perpendicular anisotropy ($\mu_0H_{2\perp}$) fields obtained from fitting are summarised in the Table.~\ref{tab:freq}. Also the values of effective saturation magnetisation field ($ M_\textrm{eff} =  M_\textrm{s} -H_{2\perp}$ ) are given in Table.~\ref{tab:freq}. It can be seen that the uniaxial anisotropy component is dominant in comparison to biaxial for all the samples. The polycrystalline samples show very little anisotropy compared to epitaxially grown samples. There can be contribution from strain induced anisotropy in epitaxially grown samples \cite{krysztofik2021effect}. The multi-layer samples exhibit large perpendicular anisotropy field ($\mu_0H_{2\perp}$) which can be explained by the scenario that the multi-layering improves the magnitude of crystal orientation and the resultant sharp interface may lead to the increase in the strain effect. As another check we also estimated the value of anisotropy fields by fitting the angular dependence of resonance field using kittel resonance formula \cite{farle1998ferromagnetic} as shown in Fig.~\ref{fig:freq} (c) and the values are summarised in Table.~\ref{tab:angle} in Appendix. The values match fairly well with those calculated from frequency dependence. 
We obtained the values of the inhomogeneous broadening, $\mu_0\Delta H_0$ and intrinsic gilbert damping, $\alpha$ for our films by fitting  the frequency dependence of linewidth, $\mu_0\Delta H$ using Eq.~\ref{eq:line}. The values obtained from fitting are summarised in Table.~\ref{tab:freq}.
\begin{align}
\mu_0\Delta H = \mu_0\Delta H_0 + \frac{2\pi\alpha}{\gamma}f
\label{eq:line}
\end{align}
 The value of $\alpha$ is fairly constant and very large $\sim$ 0.1 for all the samples as can be seen in Table.~\ref{tab:freq}. Similar damping values amongst all the samples suggest that damping is not affected by crystal structure. The enhancement of $\alpha$ in the measured films can originate from several mechanism and one such mechanism can be due to spin pumping\cite{PhysRevB.101.060402}. To quantify spin pumping term detailed analysis is required which is beyond the scope of this work. Nevertheless from Fig.~\ref{fig:freq} (a) it can be seen that the inhomogeneous contribution for samples grown on sapphire is larger than that grown on SiO$_2$ substrate which can be due to strain induced magnetic inhomogeneity. Strain can be induced by growth \cite{zhao2021growth} due to the lattice mismatch \cite{krysztofik2021effect}. In our films  there is a lattice mismatch between Pt (lattice constant $a$ = 3.93\AA) grown on sapphire (lattice constant $a$ = 4.75\AA ~\cite{inbook}) which can cause strain induced anisotropy \cite{krysztofik2021effect}.

\begin{table*}
\centering
 \caption{\label{tab:freq} Summary of the anisotropy constants and magnetisation obtained from fitting of frequency dependent resonance field and linewidth data. Note - The uncertainty values reported here are the standard error of the fitting parameters obtained from curve fitting.}

\begin{tabular}{ccccccc}
\hline\hline
Sample&  $\mu_0H_{2\parallel}$  &$\mu_0H_{2\perp}$ & $\mu_0M_{\textrm{eff}}$ &$\mu_0M_{\textrm{s}}$ &$\alpha$ &$\mu_0\Delta H_0$ \\
&(mT)  &(T) &(T) &(T) & &(T)\\
\hline
K-01 &5.62 $\pm$ 0.1  &-0.24 $\pm$ 0.006 &0.60 $\pm$ 0.06 &0.362 $\pm$ 0.05 &0.11 $\pm$ 0.01 &0.0439 $\pm$ 0.0005\\
K-02 &3.15 $\pm$ 0.01 &-0.38 $\pm$ 0.005 &0.71 $\pm$ 0.05 &0.331 $\pm$ 0.05 &0.093 $\pm$ 0.005 &0.0092 $\pm$ 0.0004\\
K-03 &-8.15 $\pm$ 0.2  &0.15 $\pm$ 0.004 &0.23 $\pm$ 0.03 &0.377 $\pm$ 0.02 &0.115 $\pm$ 0.005 &0.065 $\pm$ 0.001\\
K-04 &-2.15 $\pm$ 0.04 &-0.02 $\pm$ 0.004 &0.32 $\pm$ 0.04 &0.298 $\pm$ 0.04 &0.096 $\pm$ 0.003 &0.0093 $\pm$ 0.0007\\
\hline
\hline
\end{tabular}
\end{table*}
\begin{table*}
\centering
 \caption{\label{tab:angle} Summary of the anisotropy constants and magnetisation obtained from fitting of angular dependence of resonance field. Note - The uncertainty values reported here are the standard error of the fitting parameters obtained from curve fitting.}
\begin{tabular}{ccccc}
\hline\hline
Sample&  $\mu_0H_{2\parallel}$ &$\mu_0H_{2\perp}$ & $\mu_0M_{\textrm{eff}}$ &$\mu_0M_{\textrm{s}}$ \\
&(mT)  &(T) &(T) &(T)\\
\hline
K-01 &19.1 $\pm$ 0.1  &-0.35 $\pm$ 0.01 &0.712 $\pm$ 0.06 &0.362 $\pm$ 0.05\\
K-02 &3.01 $\pm$ 0.01  &-0.43 $\pm$ 0.01 &0.761 $\pm$ 0.05 &0.331 $\pm$ 0.05\\
K-03 &-8.1 $\pm$ 0.1 &0.142 $\pm$ 0.002 &0.235 $\pm$ 0.02 &0.377 $\pm$ 0.02\\
K-04 &-4.2 $\pm$ 0.1 &-0.011 $\pm$ 0.004 &0.309 $\pm$ 0.04 &0.298 $\pm$ 0.04\\
\hline
\hline
\end{tabular}
\end{table*}

\begin{table*}
\centering
 \caption{\label{tab:sot}SOT effective fields for different samples measured at $f=$8 GHz. All of them have been scaled by a current density of $j =10^{10}A/m^2$.}
\begin{tabular}{ccccc}
\hline\hline
Parameter &  &sample  &  & \\
 &K-01 &K-02 &K-03 &K-04\\
\hline
$\mu_0h_{\textrm y}$ (mT) &-0.14 $\pm$ 0.004 &-0.08 $\pm$ 0.007 &(1.4 $\pm$ 0.1)$\times$10$^{-2}$  &(2.7 $\pm$ 0.5)$\times$ 10$^{-3}$ \\
$\mu_0h_{\textrm x}$ (mT) &(-6.1 $\pm$ 0.2)$\times$ 10$^{-3}$ &(-5.2 $\pm$ 0.4)$\times$ 10$^{-5}$  &(-5.4 $\pm$ 0.1)$\times$ 10$^{-4}$  &(-2.5 $\pm$ 0.5)$\times$ 10$^{-4}$ \\
$\mu_0$a (mT) &(2.2 $\pm$ 0.1)$\times$ 10$^{-3}$ &(-3.2 $\pm$ 0.2)$\times$ 10$^{-3}$  &(2.4 $\pm$ 0.2)$\times$ 10$^{-3}$  &(5.6 $\pm$ 0.1)$\times$ 10$^{-3}$ \\
$\mu_0$b (mT) &-0.29 $\pm$ 0.002 &-0.245 $\pm$ 0.003 &-0.163 $\pm$ 0.002 &-0.121 $\pm$ 0.002\\
$\mu_0$c (mT) &(-9.02 $\pm$ 0.1)$\times$ 10$^{-3}$  &(-6.9 $\pm$ 0.3)$\times$ 10$^{-3}$  &(8.3 $\pm$ 0.3)$\times$ 10$^{-3}$  &(2.7 $\pm$ 0.2)$\times$ 10$^{-3}$ \\
$I_0$ (mA) &3.2 $\pm$ 0.01 &1.7 $\pm$ 0.01 &1.37 $\pm$ 0.003 &0.429 $\pm$ 0.001\\
$\Delta R (\Omega)$ &4.694 $\pm$ 0.001 &4.821 $\pm$ 0.001 &10.751 $\pm$ 0.002 &11.062 $\pm$ 0.002\\
\hline
\hline
\end{tabular}
\end{table*}

 \section{Power dependence of rectified voltage}
\begin{figure}
\centering
{\includegraphics[width=0.45\textwidth]{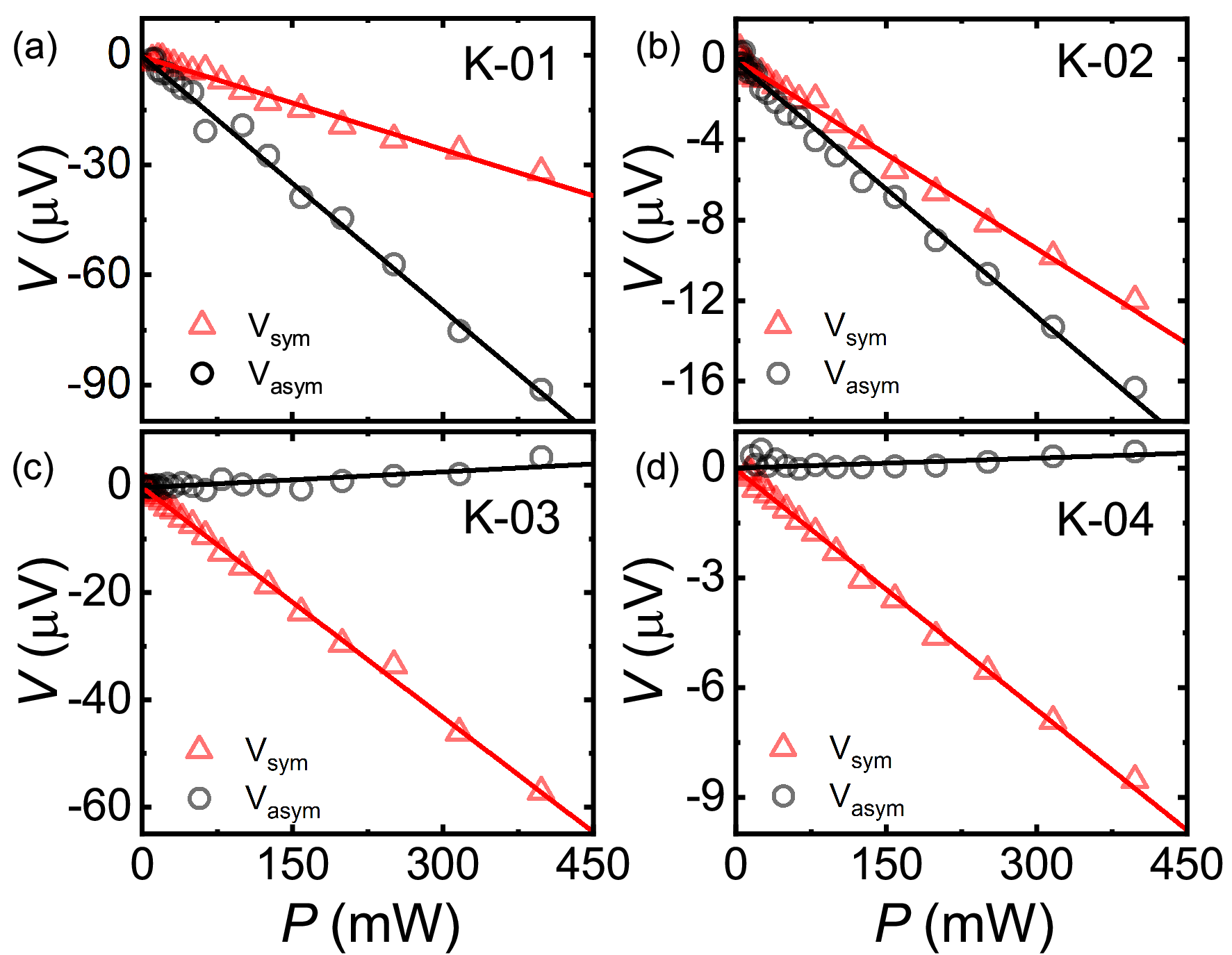}}
\caption{\label{fig:power}Plot of
magnitude of the voltage for the symmetric and anti-symmetric components as a function of injected microwave powers for different set of samples at $f$ = 8 GHz for K-01 and K-02 and 6 GHz for K-03 , K-04. All the samples show linear dependence for the given regime of injected power.}
\end{figure}

 From Eq.~\ref{eq:F13} and \ref{eq:F14} it is found that $V_{\textrm{sym}},V_{\textrm{asy}}\propto I^2$ where $h_{\textrm i}, (i=x,y,z) \propto I$. This can be seen from the linear dependence of rectified voltage $V$ on microwave power as shown in Fig.~\ref{fig:power} which is consistent with our model. Also in our experiments, the  wavelength of applied microwaves in the frequency range 3-12 GHz was much larger than the length of device ( few hundreds of $\mu$m). This ensures that current flow is uniform and the phase is almost constant \cite{li2019simultaneous}.

\section{Additional SOT-FMR results in this study}
\begin{figure}
\centering
{\includegraphics[width=0.47\textwidth]{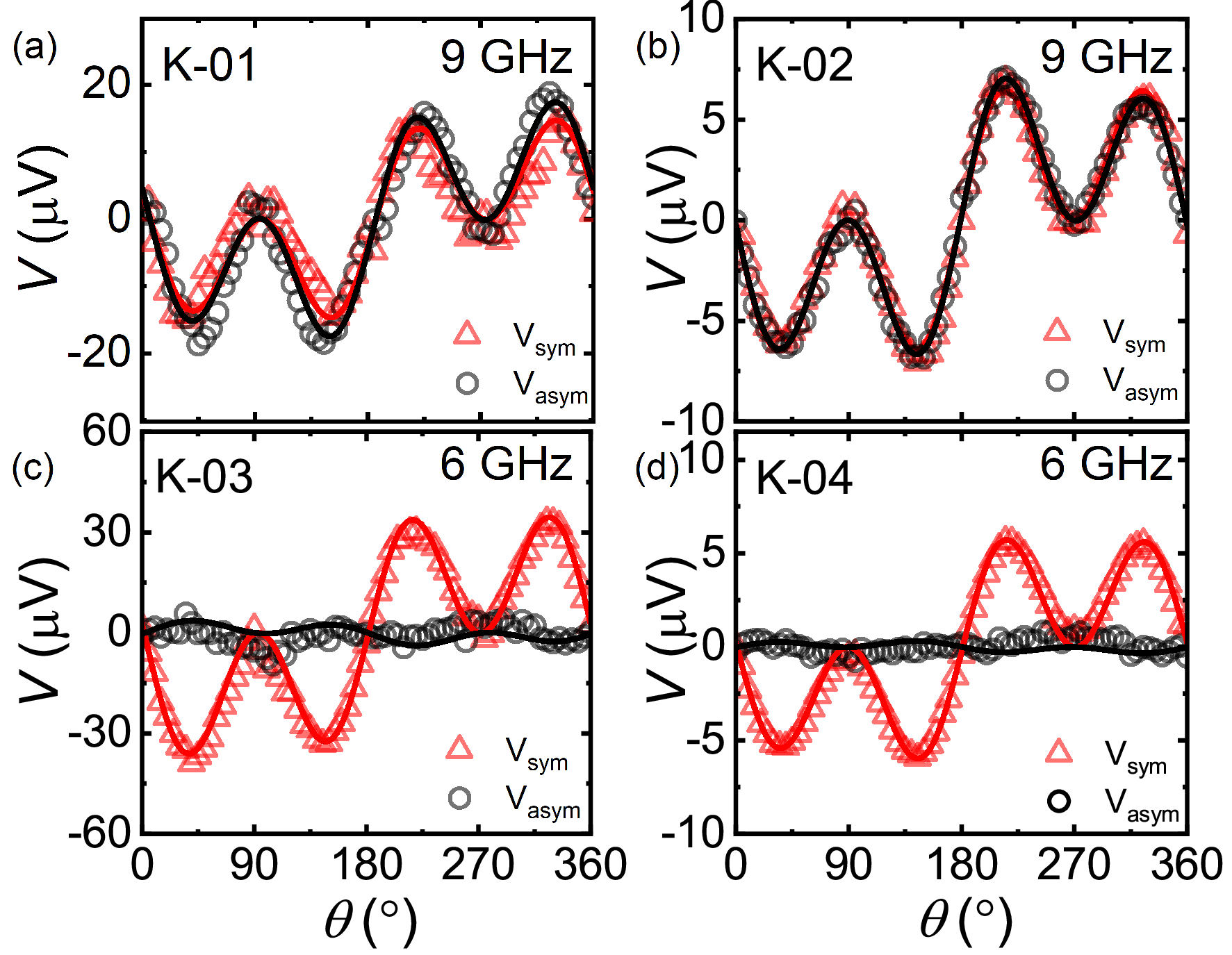}}
\caption{\label{fig:ch3_sot2}The symmetric and anti-symmetric components of the SOT-FMR spectra as a function of in-plane magnetic field angle $\theta$ for (a) K-01 (b) K-02 (c) K-03 and (d) K-04 films at different frequencies. The bi-layer samples show negligible anti-symmetric component while the anti-symmetric component is comparable to symmetric component for the multi-layer samples.}
\end{figure}
In addition to the results presented in the main text, we also investigated the angular dependence of the voltage at different frequencies as shown in Fig.~\ref{fig:ch3_sot2}. The $\theta$ dependence for samples K-01 and K-02 as shown in Fig.~\ref{fig:ch3_sot2} (a) and (b) were performed at frequency $f$ = 9 GHz while for K-3 and K-04 in bottom panel shows angular dependence performed at $f$ = 6 GHz. It can be seen that for all samples similar angular dependence as in Fig.~\ref{fig:ch3_sot} was seen regardless of the excitation frequency.

\section{Calculation of spin Hall conductivity}
The value of spin Hall conductivity ($\sigma_\textrm{SH}$) is given by the equation as below:
\begin{align}
\sigma_\textrm{SH} &= \frac{\hbar}{2e}\cdot\frac{\theta_\textrm{SHA}}{(1+\theta_\textrm{SHA}^2)\rho_\textrm{xx}}
\label{eq:cond}
\end{align}
 where, $\rho_\textrm{xx}$ is the longitudinal resistivity of Pt layer. Using Eq.~\ref{eq:cond} and the values of $\theta_\textrm{SHA}$, the value of $\sigma_\textrm{SH}$ was calculated and summarised in Table.~\ref{tab:sot2}. 
\section{Power dependence of resonance field  }
\begin{figure}
\centering
{\includegraphics[width=0.45\textwidth]{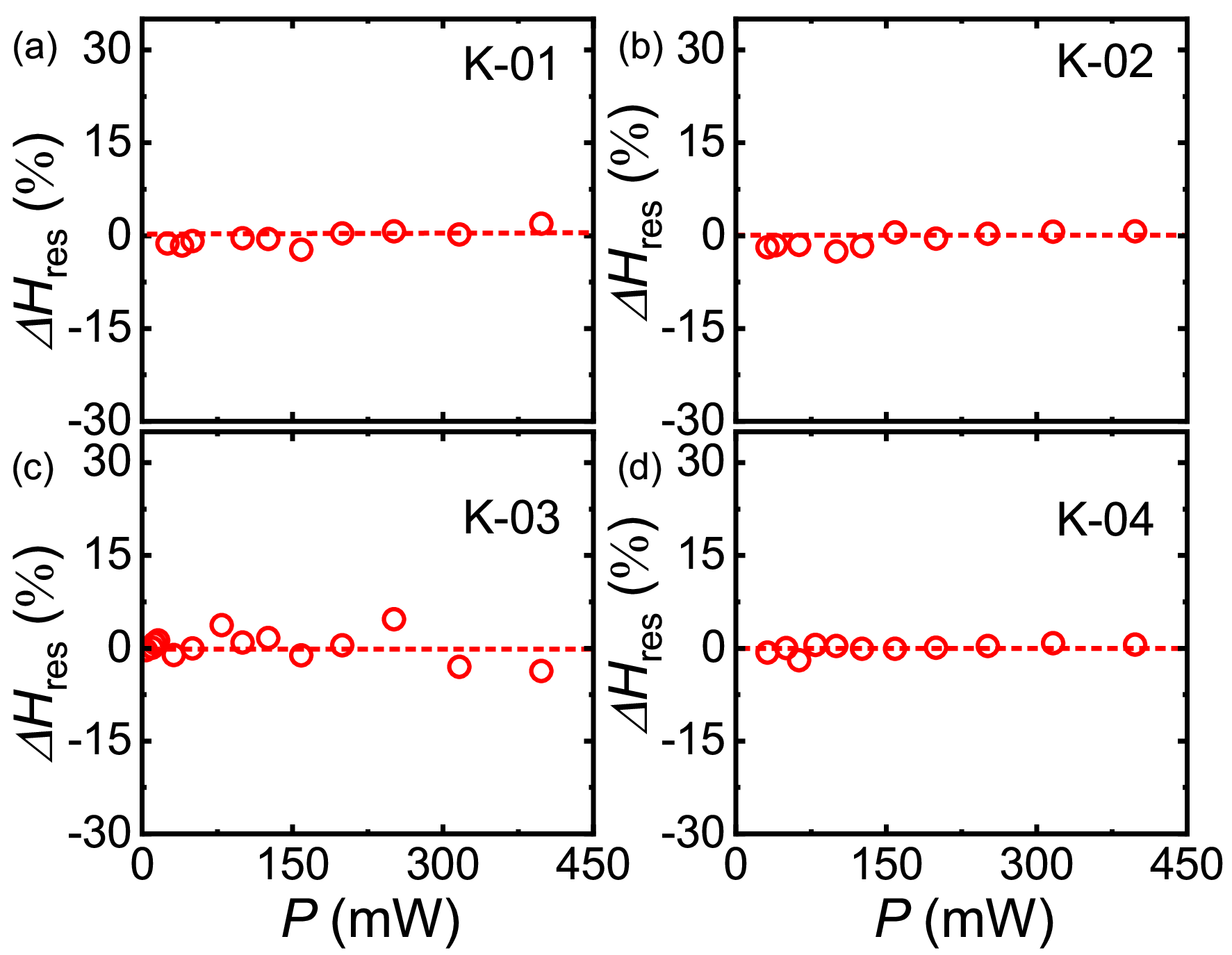}}
\caption{\label{fig:ch3_hr_fmr}Plot of
change in ratio of FMR field ($\Delta H_{\textrm{res}} = (H_{\textrm{res}}(P)- H_{\textrm{avg res}})/ H_{\textrm{avg res}}$) as a function of injected microwave powers for different set of samples at $f$ = 8 GHz for K-01 and K-02 and 6 GHz for K-03 , K-04. All the samples show zero deviation indicating that $H_{\textrm{FMR}}$ is independent of applied microwave powerthus indicating that for the given regime of input power, sample heating from power absorption does not affect the magnetic properties.}
\end{figure} 

To quantify the effect of sample heating on magnetic properties we plotted the change in resonance field, $\Delta H_{\textrm{res}}$ in Fig.~\ref{fig:ch3_hr_fmr}. It can be seen that there is no change in resonance field for the given regime of injected input microwave power. This shows that sample magnetic properties are not affected due to heating.

\normalem

\bibliographystyle{apsrev4-2}

\providecommand{\noopsort}[1]{}\providecommand{\singleletter}[1]{#1}%

\end{document}